# Combinators and the Story of Computation

Stephen Wolfram*

*We discuss the role of combinators in the development of the modern conception of computation over the course of the past century. We describe how ideas about formalism and mathematical logic led to the introduction of combinators in 1920 as an extension of the discovery of Nand as a basis for basic logic. We then discuss how combinators informed lambda calculus and symbolic computation, and their relationship to the development of practical computation. We finally describe recent views of combinators in terms of the computational universe of possible programs, and a recent approach to the fundamental theory of physics.*

## The Abstract Representation of Things

"In principle you could use combinators," some footnote might say. But the implication tends to be "But you probably don't want to." And, yes, combinators are deeply abstract—and in many ways hard to understand. But tracing their history over the hundred years since they were invented, I've come to realize just how critical they've actually been to the development of our modern conception of computation—and indeed my own contributions to it.

The idea of representing things in a formal, symbolic way has a long history. In antiquity there was Aristotle's logic and Euclid's geometry. By the 1400s there was algebra, and in the 1840s Boolean algebra. Each of these was a formal system that allowed one to make deductions purely within the system. But each, in a sense, ultimately viewed itself as being set up to model something specific. Logic was for modeling the structure of arguments, Euclid's geometry the properties of space, algebra the properties of numbers; Boolean algebra aspired to model the "laws of thought".

But was there perhaps some more general and fundamental infrastructure: some kind of abstract system that could ultimately model or represent anything? Today we understand that's what computation is. And it's becoming clear that the modern conception of computation is one of the single most powerful ideas in all of intellectual history—whose implications are only just beginning to unfold.

But how did we finally get to it? Combinators had an important role to play, woven into a complex tapestry of ideas stretching across more than a century.

A version of this document with immediately executable code is available at
https://writings.stephenwolfram.com/2020/12/combinators-and-the-story-of-computation/

* Email: s.wolfram@wolfram.com

The main part of the story begins in the 1800s. Through the course of the 1700s and 1800s mathematics had developed a more and more elaborate formal structure that seemed to be reaching ever further. But what really was mathematics? Was it a formal way of describing the world, or was it something else—perhaps something that could exist without any reference to the world?

Developments like non-Euclidean geometry, group theory and transfinite numbers made it seem as if meaningful mathematics could indeed be done just by positing abstract axioms from scratch and then following a process of deduction. But could all of mathematics actually just be a story of deduction, perhaps even ultimately derivable from something seemingly lower level—like logic?

But if so, what would things like numbers and arithmetic be? Somehow they would have to be "constructed out of pure logic". Today we would recognize these efforts as "writing programs" for numbers and arithmetic in a "machine code" based on certain "instructions of logic". But back then, everything about this and the ideas around it had to be invented.

## What Is Mathematics—and Logic—Made Of?

Before one could really dig into the idea of "building mathematics from logic" one had to have ways to "write mathematics" and "write logic". At first, everything was just words and ordinary language. But by the end of the 1600s mathematical notation like +, =, > had been established. For a while new concepts—like Boolean algebra—tended to just piggyback on existing notation. By the end of the 1800s, however, there was a clear need to extend and generalize how one wrote mathematics.

In addition to algebraic variables like $x$, there was the notion of symbolic functions $f$, as in $f(x)$. In logic, there had long been the idea of letters ($p$, $q$, ...) standing for propositions ("it is raining now"). But now there needed to be notation for quantifiers ("for all $x$ such-and-such", or "there exists $x$ such that ..."). In addition, in analogy to symbolic functions in mathematics, there were symbolic logical predicates: not just explicit statements like $x > y$ but also ones like $p(x, y)$ for symbolic $p$.

The first full effort to set up the necessary notation and come up with an actual scheme for constructing arithmetic from logic was Gottlob Frege's 1879 *Begriffsschrift* ("concept script"):

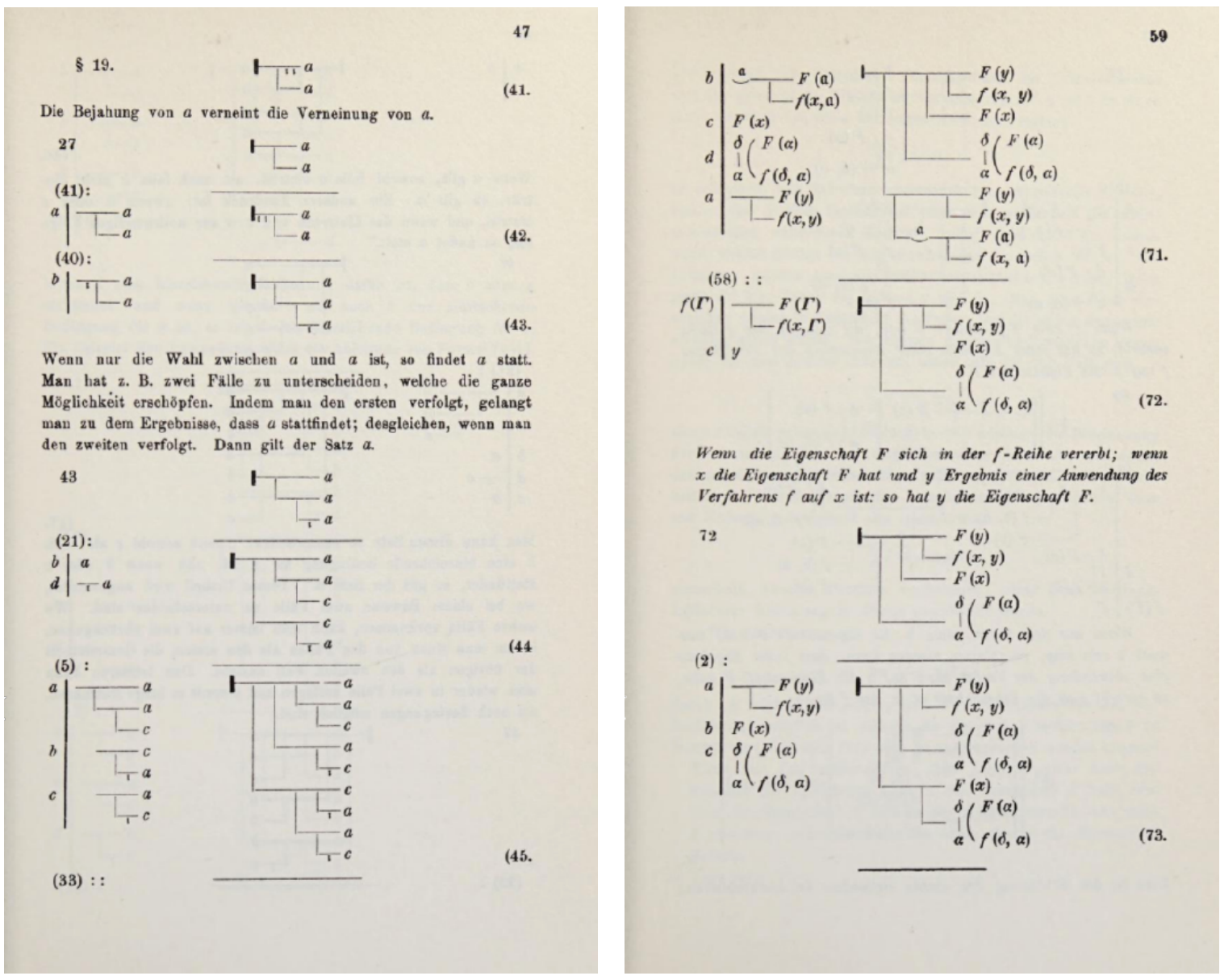
Wenn nur die Wahl zwischen *a* und *a* ist, so findet *a* statt. Man hat z. B. zwei Fälle zu unterscheiden, welche die ganze Möglichkeit erschöpfen. Indem man den ersten verfolgt, gelangt man zu dem Ergebnisse, dass *a* stattfindet; desgleichen, wenn man den zweiten verfolgt. Dann gilt der Satz *a*.

*Wenn die Eigenschaft F sich in der f-Reihe vererbt; wenn x die Eigenschaft F hat und y Ergebnis einer Anwendung des Verfahrens f auf x ist: so hat y die Eigenschaft F.*

And, yes, it was not so easy to read, or to typeset—and at first it didn't make much of an impression. But the notation got more streamlined with Giuseppe Peano's *Formulario* project in the 1890s—which wasn't so concerned with starting from logic as starting from some specified set of axioms (the "Peano axioms"):

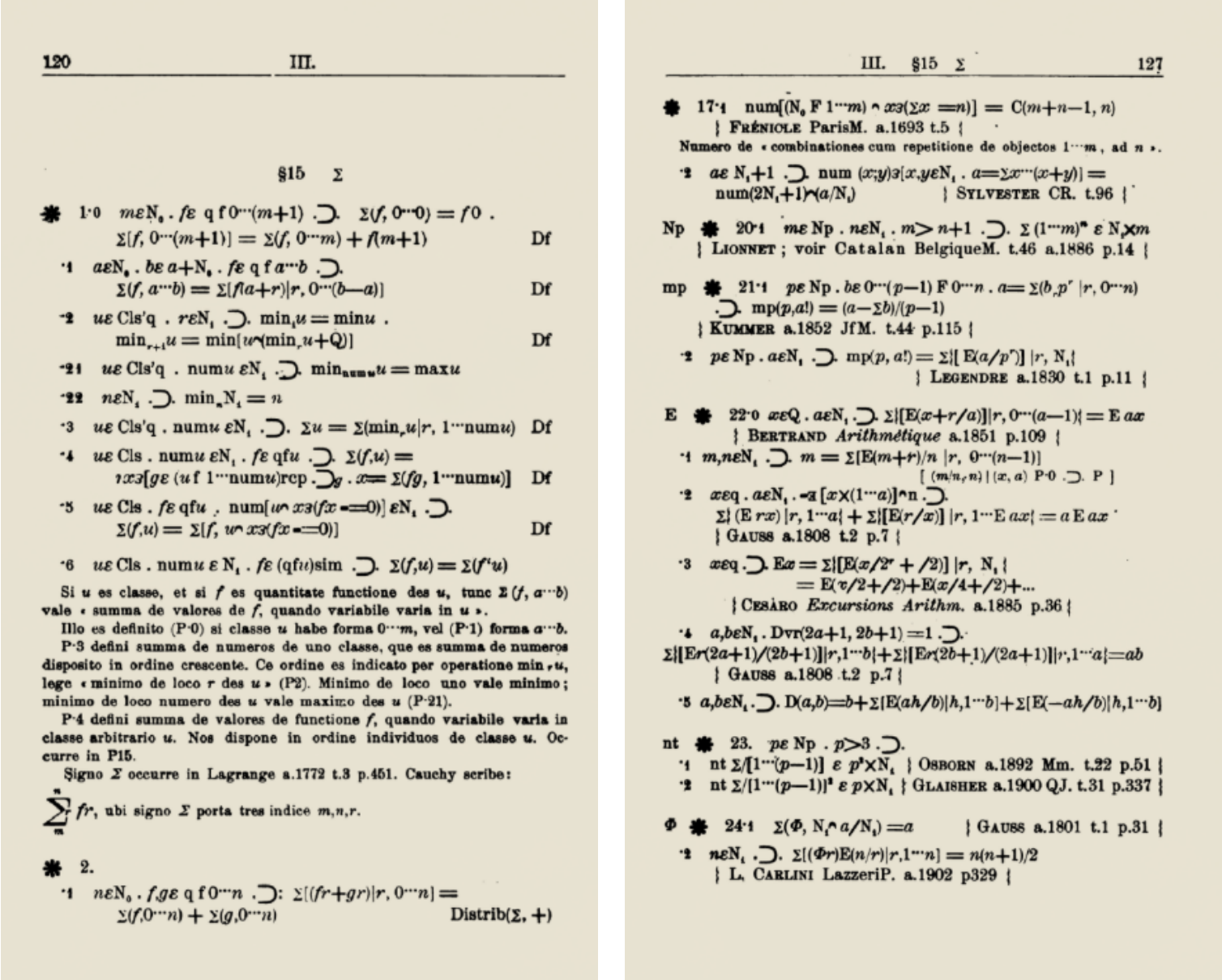
Si *u* es classe, et si *f* es quantitate functione des *u*, tunc Σ (*f*, *a*···*b*) vale « summa de valores de *f*, quando variabile varia in *u* ».
Illo es definito (P·0) si classe *u* habe forma 0···*m*, vel (P·1) forma *a*···*b*.
P·3 defini summa de numeros de uno classe, que es summa de numeros disposito in ordine crescente. Ce ordine es indicato per operatione min$_r$*u*, lege « minimo de loco *r* des *u* » (P2). Minimo de loco uno vale minimo; minimo de loco numero des *u* vale maximo des *u* (P·21).
P·4 defini summa de valores de functione *f*, quando variabile varia in classe arbitrario *u*. Nos dispone in ordine individuos de classe *u*. Occurre in P15.
Signo Σ occurre in Lagrange a.1772 t.3 p.451. Cauchy scribe: $\sum_m^n fr$, ubi signo Σ porta tres indice *m*,*n*,*r*.

And then in 1910 Alfred Whitehead and Bertrand Russell began publishing their 2000-page *Principia Mathematica*—which pretty much by its sheer weight and ambition (and notwithstanding what I would today consider grotesque errors of language design)—popularized the possibility of building up "the complexity of mathematics" from "the simplicity of logic":

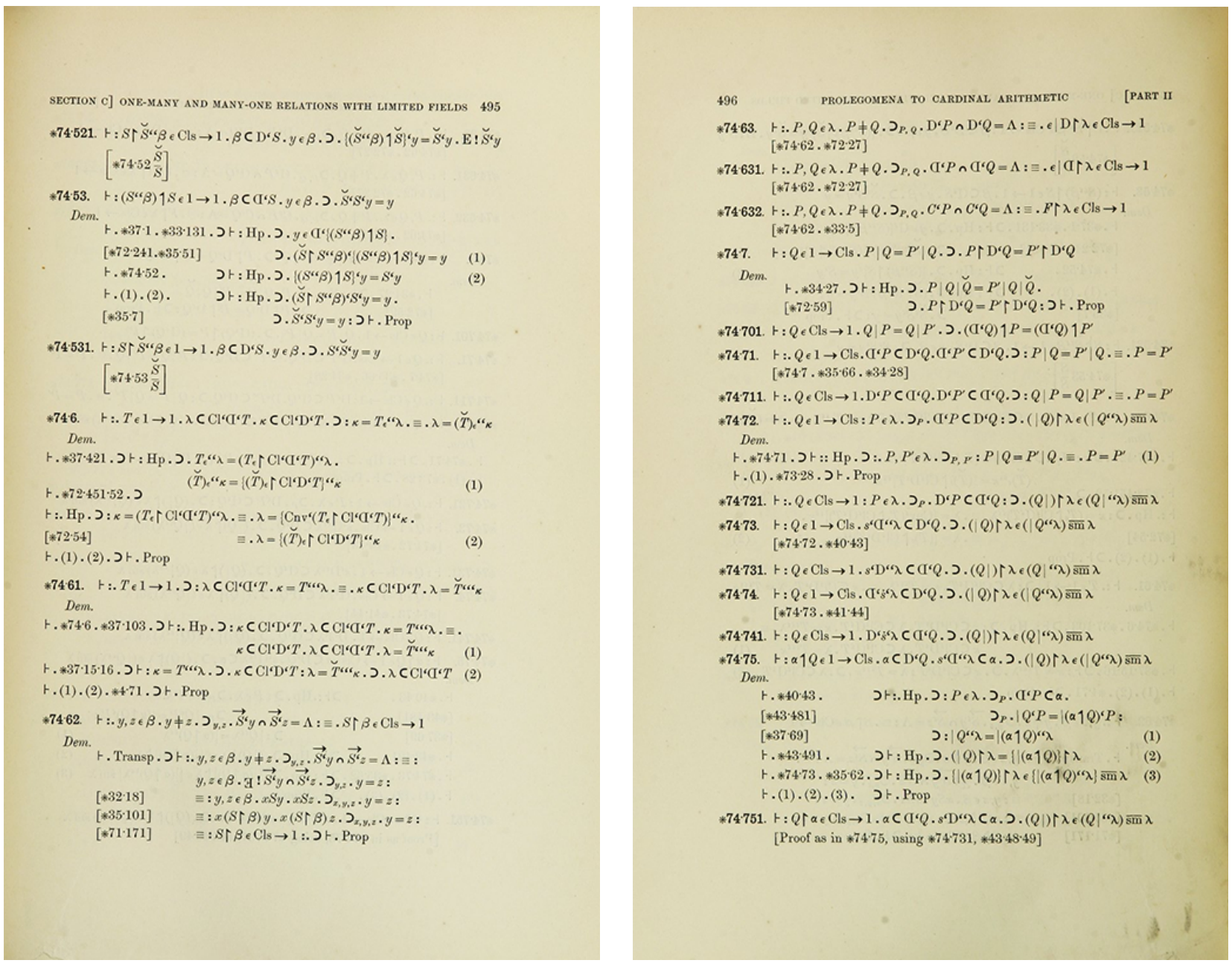
SECTION C] ONE-MANY AND MANY-ONE RELATIONS WITH LIMITED FIELDS 495

496 PROLEGOMENA TO CARDINAL ARITHMETIC [PART II

It was one thing to try to represent the content of mathematics, but there was also the question of representing the infrastructure and processes of mathematics. Let's say one picks some axioms. How can one know if they're consistent? What's involved in proving everything one can prove from them?

In the 1890s David Hilbert began to develop ideas about this, particularly in the context of tightening up the formalism of Euclid's geometry and its axioms. And after *Principia Mathematica*, Hilbert turned more seriously to the use of logic-based ideas to develop "metamathematics"—notably leading to the formulation of things like the "decision problem" (*Entscheidungsproblem*) of asking whether, given an axiom system, there's a definite procedure to prove or disprove any statement with respect to it.

But while connections between logic and mathematics were of great interest to people concerned with the philosophy of mathematics, a more obviously mathematical development was universal algebra—in which axioms for different areas of mathematics were specified just by giving appropriate algebraic-like relations. (As it happens, universal algebra was launched under that name by the 1898 book *A Treatise on Universal Algebra* by Alfred Whitehead, later of *Principia Mathematica* fame.)

But there was one area where ideas about algebra and logic intersected: the tightening up of Boolean algebra, and in particular the finding of simpler foundations for it. Logic had pretty much always been formulated in terms of **And**, **Or** and **Not**. But in 1912 Henry Sheffer—attempting to simplify *Principia Mathematica*—showed that just **Nand** (or **Nor**) were sufficient. (It turned out that Charles Peirce had already noted the same thing in the 1880s.)

So that established that the notation of logic could be made basically as simple as one could imagine. But what about its actual structure, and axioms? Sheffer talked about needing five "algebra-style" axioms. But by going to axioms based on logical inferences Jean Nicod managed in 1917 to get it down to just one axiom. (And, as it happens, I finally finished the job in 2000 by finding the very simplest "algebra-style" axioms for logic—the single axiom: $((p\cdot q)\cdot r)\cdot(p\cdot((p\cdot r)\cdot p))=r$.)

The big question had in a sense been "What is mathematics ultimately made of?". Well, now it was known that ordinary propositional logic could be built up from very simple elements. So what about the other things used in mathematics—like functions and predicates? Was there a simple way of building these up too?

People like Frege, Whitehead and Russell had all been concerned with constructing specific things—like sets or numbers—that would have immediate mathematical meaning. But Hilbert's work in the late 1910s began to highlight the idea of looking instead at metamathematics and the "mechanism of mathematics"—and in effect at how the pure symbolic infrastructure of mathematics fits together (through proofs, etc.), independent of any immediate "external" mathematical meaning.

Much as Aristotle and subsequent logicians had used (propositional) logic to define a "symbolic structure" for arguments, independent of their subject matter, so too did Hilbert's program imagine a general "symbolic structure" for mathematics, independent of particular mathematical subject matter.

And this is what finally set the stage for the invention of combinators.

## Combinators Arrive

We don't know how long it took Moses Schönfinkel to come up with combinators. From what we know of his personal history, it could have been as long as a decade. But it could also have been as short as a few weeks.

There's no advanced math or advanced logic involved in defining combinators. But to drill through the layers of technical detail of mathematical logic to realize that it's even conceivable that everything can be defined in terms of them is a supreme achievement of a kind of abstract reductionism.

There is much we don't know about Schönfinkel as a person. But the 11-page paper he wrote on the basis of his December 7, 1920, talk in which he introduced combinators is extremely clear.

The paper is entitled "On the Building Blocks of Mathematical Logic" (in the original German, "Über die Bausteine der mathematischen Logik".) In other words, its goal is to talk about "atoms" from which mathematical logic can be built. Schönfinkel explains that it's "in the spirit of" Hilbert's axiomatic method to build everything from as few notions as possible; then he says that what he wants to do is to "seek out those notions from which we shall best be able to construct all other notions of the branch of science in question".

His first step is to explain that Hilbert, Whitehead, Russell and Frege all set up mathematical logic in terms of standard And, Or, Not, etc. connectives—but that Sheffer had recently been able to show that just a single connective (indicated by a stroke "|"—and what we would now call Nand) was sufficient:

*a* und *b* ist mindestens eine falsch", die mit den obigen Zeichen in den beiden äquivalenten Formen

$$\bar{a} \vee \bar{b} \quad \text{und} \quad \overline{a \,\&\, b}$$

geschrieben werden kann, und als zugehöriges neues Zeichen

$$a \mid b,$$

so ist augenscheinlich

$$\bar{a} = a \mid a, \qquad a \vee b = (a \mid a) \mid (b \mid b),$$

womit wegen

$$a \,\&\, b = \overline{\bar{a} \vee \bar{b}}, \qquad (a \rightarrow b) = \bar{a} \vee b, \qquad (a \sim b) = (a \rightarrow b) \,\&\, (b \rightarrow a)$$

But in addition to the "content" of these relations, I think Schönfinkel was trying to communicate by example something else: that all these logical connectives can ultimately be thought of just as examples of "abstract symbolic structures" with a certain "function of arguments" (i.e. *f*[*x*,*y*]) form.

The next couple of paragraphs talk about how the quantifiers "for all" (∀) and "there exists" (∃) can also be simplified in terms of the Sheffer stroke (i.e. Nand). But then comes the rallying cry: "The successes that we have encountered thus far ... encourage us to attempt further progress." And then he's ready for the big idea—which he explains "at first glance certainly appears extremely bold". He proposes to "eliminate by suitable reduction the remaining fundamental concepts of proposition, function and variable".

He explains that this only makes sense for "arbitrary, logically general propositions", or, as we'd say now, for purely symbolic constructs without specific meanings yet assigned. In other words, his goal is to create a general framework for operating on arbitrary symbolic expressions independent of their interpretation.

He explains that this is valuable both from a "methodological point of view" in achieving "the greatest possible conceptual uniformity", but also from a certain philosophical or perhaps aesthetic point of view.

And in a sense what he was explaining—back in 1920—was something that's been a core part of the computational language design that I've done for the past 40 years: that everything can be represented as a symbolic expression, and that there's tremendous value to this kind of uniformity.

But as a “language designer” Schönfinkel was an ultimate minimalist. He wanted to get rid of as many notions as possible—and in particular he didn’t want variables, which he explained were “nothing but tokens that characterize certain argument places and operators as belonging together”; “mere auxiliary notions”.

Today we have all sorts of mathematical notation that’s at least somewhat “variable free” (think coordinate-free notation, category theory, etc.) But in 1920 mathematics as it was written was full of variables. And it needed a serious idea to see how to get rid of them. And that’s where Schönfinkel starts to go “even more symbolic”.

He explains that he’s going to make a kind of “functional calculus” (*Funktionalkalkül*). He says that normally functions just define a certain correspondence between the domain of their arguments, and the domain of their values. But he says he’s going to generalize that—and allow (“disembodied”) functions to appear as arguments and values of functions. In other words, he’s inventing what we’d now call higher-order functions, where functions can operate “symbolically” on other functions.

In the context of traditional calculus-and-algebra-style mathematics it’s a bizarre idea. But really it’s an idea about computation and computational structures—that’s more abstract and ultimately much more general than the mathematical objectives that inspired it.

But back to Schönfinkel’s paper. His next step is to explain that once functions can have other functions as arguments, functions only ever need to take a single argument. In modern (Wolfram Language) notation he says that you never need f[x,y]; you can always do everything with f[x][y].

In something of a sleight of hand, he sets up his notation so that *fxyz* (which might look like a function of three arguments f[x,y,z]) actually means (((*fx*)*y*)*z*) (i.e. f[x][y][z]). (In other words—somewhat confusingly with respect to modern standard functional notation—he takes function application to be left associative.)

Again, it’s a bizarre idea—though actually Frege had had a similar idea many years earlier (and now the idea is usually called currying, after Haskell Curry, who we’ll be talking about later). But with his “functional calculus” set up, and all functions needing to take only one argument, Schönfinkel is ready for his big result.

He’s effectively going to argue that by combining a small set of particular functions he can construct any possible symbolic function—or at least anything needed for predicate logic. He calls them a “sequence of particular functions of a very general nature”. Initially there are five of them: the identity function (*Identitätsfunktion*) *I*, the constancy function (*Konstanzfunktion*) *C* (which we now call K), the interchange function (*Vertauschungsfunktion*) *T*, the composition function (*Zusammensetzungsfunktion*) *Z*, and the fusion function (*Verschmelzungsfunktion*) *S*.

§ 3.

Es soll nunmehr eine Reihe von *individuellen Funktionen* von sehr allgemeiner Natur eingeführt werden. Ich nenne sie: die Identitätsfunktion $I$, die Konstanzfunktion $C$, die Vertauschungsfunktion $T$, die Zusammensetzungsfunktion $Z$ und die Verschmelzungsfunktion $S$.

1. Unter der *Identitätsfunktion* $I$ soll diejenige völlig bestimmte Funktion verstanden werden, deren Argumentwert keiner Einschränkung unterworfen ist und deren Funktionswert stets mit dem Argumentwert übereinstimmt, durch die also jedes Ding und jede Funktion sich selbst zugeordnet wird. Sie ist somit definiert durch die Gleichung

$$I x = x,$$

in welcher das Gleichheitszeichen nicht etwa als logische Äquivalenz im Sinne der im logischen Aussagenkalkül üblichen Definition zu lesen ist, sondern besagt, daß die Ausdrücke links und rechts dasselbe bedeuten, d. h. daß der Funktionswert $Ix$ stets derselbe ist wie der Argumentwert $x$, was man auch für $x$ einsetzen mag. (So wäre z. B. $II = I$.)

2. Nunmehr sei der Argumentwert wieder ohne Einschränkung beliebig, während der Funktionswert unabhängig von jenem stets der feste Wert $a$ sein soll. Diese Funktion ist ihrerseits von $a$ abhängig, also von der Form $Ca$. Daß ihr Funktionswert stets $a$ ist, wird geschrieben:

$$(C a) y = a.$$

Und, indem wir nun auch $a$ variabel lassen, erhalten wir:

$$(C x) y = x \quad \text{bzw.} \quad C x y = x$$

als Definitionsgleichung der *Konstanzfunktion* $C$. Diese Funktion $C$ ist augenscheinlich von der auf S. 308 betrachteten Art; sie liefert nämlich erst durch Einsetzen eines festen Wertes für $x$ eine Funktion mit dem Argument $y$. In der praktischen Anwendung leistet sie uns den Dienst, daß sie eine Größe $x$ als „blinde" Veränderliche einzuführen gestattet.

And then he's off and running defining what we now call combinators. The definitions look simple and direct. But to get to them Schönfinkel effectively had to cut away all sorts of conceptual baggage that had come with the historical development of logic and mathematics.

Even talking about the identity combinator isn't completely straightforward. Schönfinkel carefully explains that in *I x* = *x*, equality is direct symbolic or structural equality, or as he puts it "the equal sign is not to be taken to represent logical equivalence as it is ordinarily defined in the propositional calculus of logic but signifies that the expressions on the left and on the right mean the same thing, that is, that the function value *lx* is always the same as the argument value *x*, whatever we may substitute for *x*." He then adds parenthetically, "Thus, for instance, *I I* would be equal to *I*". And, yes, to someone used to the mathematical idea that a function takes values like numbers, and gives back numbers, this is a bit mind-blowing.

Next he explains the constancy combinator, that he called *C* (even though the German word for it starts with *K*), and that we now call *K*. He says "let us assume that the argument value is again arbitrary without restriction, while, regardless of what this value is, the function value will always be the fixed value *a*". And when he says "arbitrary" he really means it:

it's not just a number or something; it's what we would now think of as any symbolic expression.

First he writes $(C\ a)y = a$, i.e. the value of the "constancy function $C\ a$ operating on any $y$ is $a$", then he says to "let a be variable too", and defines $(C\ x)y = x$ or $Cxy = x$. Helpfully, almost as if he were writing computer documentation, he adds: "In practical applications $C$ serves to permit the introduction of a quantity $x$ as a 'blind' variable."

Then he's on to $T$. In modern notation the definition is $T[f][x][y] = f[y][x]$ (i.e. $T$ is essentially ReverseApplied). (He wrote the definition as $(T\phi)xy{=}\phi yx$, explaining that the parentheses can be omitted.) He justifies the idea of $T$ by saying that "The function $T$ makes it possible to alter the order of the terms of an expression, and in this way it compensates to a certain extent for the lack of a commutative law."

Next comes the composition combinator $Z$. He explains that "In [mathematical] analysis, as is well known, we speak loosely of a 'function of a function'...", by which he meant that it was pretty common then (and now) to write something like $f(g(x))$. But then he "went symbolic"—and defined a composition function that could symbolically act on any two functions $f$ and $g$: $Z[f][g][x] = f[g[x]]$. He explains that $Z$ allows one to "shift parentheses" in an expression: i.e. whatever the objects in an expression might be, $Z$ allows one to transform [][][] to [[]] etc. But in case this might have seemed too abstract and symbolic, he then attempted to explain in a more "algebraic" way that the effect of $Z$ is "somewhat like that of the associative law" (though, he added, the actual associative law is not satisfied).

Finally comes the *pièce de résistance*: the S combinator (that Schönfinkel calls the "fusion function"):

5. Setzt man in

$$f\,x\,y$$

für $y$ den Wert einer Funktion $g$ ein, und zwar genommen für dasselbe $x$, das als Argument von $f$ auftritt, so kommt man auf einen Ausdruck

$$f\,x\,(g\,x)$$

oder, wie wir für den Augenblick etwas übersichtlicher schreiben wollen:

$$(f\,x)(g\,x).$$

Dies ist natürlich der Wert einer Funktion von $x$ allein, also

$$(f\,x)(g\,x) = F\,x,$$

wo

$$F = S'(f, g)$$

wieder in einer völlig bestimmten Weise von den gegebenen Funktionen $f$ und $g$ abhängt. Wir haben demgemäß:

$$[S'(\varphi, \chi)]\,x = (\varphi\,x)(\chi\,x)$$

oder, nach der auch im vorigen Fall verwendeten Umformung:

$$S\,\varphi\,\chi\,x = (\varphi\,x)(\chi\,x)$$

als Definitionsgleichung der *Verschmelzungsfunktion* $S$.

Es wird gut sein, diese Funktion durch ein praktisches Beispiel dem Verständnis näherzubringen. Nehmen wir etwa für $f\,x\,y$ den Wert ${}^{x}\log y$ (d. h. den Logarithmus von $y$ zu der Basis $x$) und für $g\,z$ den Funktionswert $1 + z$, so ergibt sich $(f\,x)(g\,x)$ augenscheinlich als ${}^{x}\log(1 + x)$, d. h. als der Wert einer Funktion von $x$, die mit den beiden gegebenen Funktionen eben durch unsere allgemeine Funktion $S$ eindeutig verknüpft ist.

Der praktische Nutzen der Funktion $S$ besteht ersichtlich darin, daß sie es ermöglicht, mehrmals auftretende Veränderliche — und bis zu einem gewissen Grade auch individuelle Funktionen — nur einmal auftreten zu lassen.

He doesn't take too long to define it. He basically says: consider $(fx)(gx)$ (i.e. f[x][g[x]]). This is really just "a function of $x$". But what function? It's not a composition of $f$ and $g$; he calls it a "fusion", and he defines the $S$ combinator to create it: $S[f][g][x] = f[x][g[x]]$.

It's pretty clear Schönfinkel knew this kind of "symbolic gymnastics" would be hard for people to understand. He continues: "It will be advisable to make this function more intelligible by means of a practical example." He says to take *fxy* (i.e. f[x][y]) to be $\log_x y$ (i.e. Log[x,y]), and *gz* (i.e. g[z]) to be $1 + z$. Then $Sfgx=(fx)\,(gx)=\log_x(1+x)$ (i.e. S[f][g][x]=f[x][g[x]]=Log[x,1+x]). And, OK, it's not obvious why one would want to do that, and I'm not rushing to make $S$ a built-in function in the Wolfram Language.

But Schönfinkel explains that for him "the practical use of the function $S$ will be to enable us to reduce the number of occurrences of a variable—and to some extent also of a particular function—from several to a single one".

Setting up everything in terms of five basic objects $I$, $C$ (now $K$), $T$, $Z$ and $S$ might already seem impressive and minimalist enough. But Schönfinkel realized that he could go even further:

Es wird sich für die Durchführung unseres logisch-symbolischen Problems als belangreich erweisen, daß die oben erklärten fünf individuellen Funktionen $I, C, T, Z, S$ des Funktionenkalküls nicht voneinander unabhängig sind, vielmehr zwei von ihnen, nämlich $C$ und $S$, hinreichen, um die übrigen durch sie zu definieren. Und zwar bestehen hier die folgenden Zusammenhänge:

1. Es ist gemäß der Erklärung der Funktionen $I$ und $C$:

$$Ix = x = Cxy.$$

Da $y$ willkürlich ist, können wir dafür ein beliebiges Ding oder eine beliebige Funktion einsetzen, also z. B. $Cx$. Dies gibt:

$$Ix = (Cx)(Cx).$$

Nach der Erklärung von $S$ bedeutet dies aber:

$$SCCx,$$

so daß wir erhalten:

$$I = SCC.$$ [3])

Übrigens kommt es in dem Ausdruck $SCC$ auf das letzte Zeichen $C$ gar nicht einmal an. Setzen wir nämlich oben für $y$ nicht $Cx$, sondern die willkürliche Funktion $\varphi x$, so ergibt sich entsprechend:

$$I = SC\varphi,$$

wo also für $\varphi$ jede beliebige Funktion eingesetzt werden kann[4]).

2. Nach der Erklärung von $Z$ ist

$$Zfgx = f(gx).$$

Weiter ist vermöge der bereits verwendeten Umformungen

$$f(gx) = (Cfx)(gx) = S(Cf)gx = (CSf)(Cf)gx.$$

Verschmelzung nach $f$ ergibt:

$$S(CS)Cfgx,$$

also

$$Z = S(CS)C.$$

3. Ganz entsprechend läßt sich

$$Tfyx = fxy$$

weiter umformen in:

$$fx(Cyx) = (fx)(Cyx) = Sf(Cy)x = (Sf)(Cy)x = Z(Sf)Cyx$$
$$= ZZSfCyx = (ZZSf)Cyx = (ZZSf)(CCf)yx = S(ZZS)(CC)fyx.$$

Es gilt somit:

$$T = S(ZZS)(CC).$$

Setzt man hier für $Z$ den oben gefundenen Ausdruck ein, so ist damit $T$ ebenfalls auf $C$ und $S$ zurückgeführt.

First, he says that actually *I* = *SCC* (or, in modern notation, s[k][k]). In other words, s[k][k][x] for symbolic *x* is just equal to *x* (since s[k][k][x] becomes k[x][k[x]] by using the definition of *S*, and this becomes *x* by using the definition of *C*). He notes that this particular reduction was communicated to him by a certain Alfred Boskowitz (who we know to have been a student at the time); he says that Paul Bernays (who was more of a colleague) had "some time before" noted that *I* = (*SC*)(*CC*) (i.e. s[k][k[k]]). Today, of course, we can use a computer to just enumerate all possible combinator expressions of a particular size, and find what the smallest reduction is. But in Schönfinkel's day, it would have been more like solving a puzzle by hand.

Schönfinkel goes on, and proves that *Z* can also be reduced: *Z* = *S*(*CS*)*C* (i.e. s[k[s]][k]). And,

yes, a very simple Wolfram Language program can verify in a few milliseconds that that is the simplest form.

OK, what about $T$? Schönfinkel gives 8 steps of reduction to prove that $T = S(ZZS)(CC)$ (i.e. s[s[k[s]][k][s[k[s]][k]][s]][k[k]]). But is this the simplest possible form for $T$? Well, no. But (with the very straightforward 2-line Wolfram Language program I wrote) it did take my modern computer a number of minutes to determine what the simplest form is.

The answer is that it doesn't have size 12, like Schönfinkel's, but rather size 9. Actually, there are 6 cases of size 9 that all work: s[s[k[s]][s[k[k]][s]]][k[k]] ($S(S(KS)(S(KK)S))(KK)$) and five others. And, yes, it takes a few steps of reduction to prove that they work (the other size-9 cases $S(SSK(K(SS(KK))))S$, $S(S(K(S(KS)K))S)(KK)$, $S(K(S(S(KS)K)(KK)))S$, $S(K(SS(KK)))(S(KK)S)$, $S(K(S(K(SS(KK)))K))S$ all have more complicated reductions):

```
s[s[k[s]][s[k[k]][s]]][k[k]][f][g][x]
s[k[s]][s[k[k]][s]][f][k[k][f]][g][x]
k[s][f][s[k[k]][s][f]][k[k][f]][g][x]
s[s[k[k]][s][f]][k[k][f]][g][x]
s[k[k]][s][f][g][k[k][f][g]][x]
k[k][f][s[f]][g][k[k][f][g]][x]
k[s[f]][g][k[k][f][g]][x]
s[f][k[k][f][g]][x]
f[x][k[k][f][g][x]]
f[x][k[g][x]]
f[x][g]
```

But, OK, what did Schönfinkel want to do with these objects he'd constructed? As the title of his paper suggests, he wanted to use them as building blocks for mathematical logic. He begins: "Let us now apply our results to a special case, that of the calculus of logic in which the basic elements are individuals and the functions are propositional functions." I consider this sentence significant. Schönfinkel didn't have a way to express it (the concept of universal computation hadn't been invented yet), but he seems to have realized that what he'd done was quite general, and went even beyond being able to represent a particular kind of logic.

Still, he went on to give his example. He'd explained at the beginning of the paper that the quantifiers we now call ∀ and ∃ could both be represented in terms of a kind of "quantified Nand" that he wrote $|^{x}$ :

Es ist nun bemerkenswerterweise sogar noch darüber hinaus möglich, durch eine geeignete Abänderung der Grundverknüpfung auch die beiden höheren Aussagen

$$(x)f(x) \quad \text{und} \quad (Ex)f(x),$$

d. h. „Alle Individuen haben die Eigenschaft $f$" und „Es gibt ein Individuum, das die Eigenschaft $f$ hat", mit andern Worten, die beiden Operationen $(x)$ und $(Ex)$, die mit den früheren zusammen bekanntlich ein im Sinne der Axiomatik vollständiges System von Grundverknüpfungen der mathematischen Logik ausmachen, mit zu erfassen.

Verwenden wir nämlich als Grundbeziehung nunmehr

$$(x)[\overline{f(x)} \vee \overline{g(x)}] \quad \text{bzw.} \quad (x)\overline{f(x) \,\&\, g(x)}$$

und schreiben wir hierfür

$$f(x) \mid^x g(x),$$

so gilt offenbar (da wir Konstante formal wie Funktionen eines Arguments behandeln dürfen):

$$\bar{a} = a \mid^x a, \qquad a \vee b = (x)(\bar{\bar{a}} \vee \bar{\bar{b}}) = \bar{a} \mid^x \bar{b} = (a \mid^y a) \mid^x (b \mid^y b),$$

$$(x)f(x) = (x)(\overline{\overline{f(x)}} \vee \overline{\overline{f(x)}}) = \overline{f(x)} \mid^x \overline{f(x)} = (f(x) \mid^y f(x)) \mid^x (f(x) \mid^y f(x)),$$

womit wegen

$$(Ex)f(x) = \overline{(x)\overline{f(x)}}$$

auch die neue Behauptung erwiesen ist.

But now he wanted to "combinator-ify" everything. So he introduced a new combinator *U*, and defined it to represent his "quantified Nand": $Ufg = fx \mid^x gx$ (he called *U* the "incompatibility function"—an interesting linguistic description of Nand):

Bausteine der mathematischen Logik. 313

die Funktionen die Aussagefunktionen sind. Wir brauchen zunächst eine weitere individuelle Funktion, die diesem Kalkül eigentümlich ist. Der Ausdruck

$$fx \mid^x gx,$$

wo $f$ und $g$ Aussagefunktionen eines Arguments sind — auf solche dürfen wir uns gemäß einer früheren Bemerkung beschränken —, ist augenscheinlich eine bestimmte Funktion der beiden Funktionen $f$ und $g$, also von der Form $U(f, g)$ oder, nach unserem Umformungsprinzip, $Ufg$. Damit haben wir

$$Ufg = fx \mid^x gx,$$

wo $f$ und $g$ nun natürlich Aussagefunktionen sind, als Definitionsgleichung der *Unverträglichkeitsfunktion* $U$.

Es besteht nun die bemerkenswerte Tatsache, daß jede logische Formel sich allein durch unsere individuellen Funktionen $I, C, T, Z, S, U$, also insbesondere schon durch $C, S$ und $U$ ausdrücken läßt.

Zunächst einmal läßt sich jede logische Formel vermittels der verallgemeinerten Strichsymbolik ausdrücken, wobei die gebundenen Veränderlichen (apparent variables) an den oberen Enden der Striche stehen. Dies gilt ohne Einschränkung, also für beliebige Aussagenordnungen und auch, wenn Beziehungen auftreten. Weiterhin läßt sich schrittweise mit geeigneter Verwendung der übrigen konstanten Funktionen an Stelle des Strichsymbols die Funktion $U$ einführen.

Der Nachweis soll hier nicht vollständig durchgeführt, sondern nur die Rolle der verschiedenen individuellen Funktionen bei dieser Zurückführung erläutert werden.

Vermöge der Funktion $C$ kann man erreichen, daß die beiden links und rechts vom Strich stehenden Ausdrücke Funktionen desselben Arguments sind.

So wäre z. B. der von $f, g$ und $y$ abhängige Ausdruck

$$fx \mid^x gy,$$

wo also rechts $x$ nicht vorkommt, als

$$fx \mid^x C(gy)x$$

umzuschreiben. Kommt dagegen $x$ rechts an anderer Stelle vor, so läßt es sich vermittels der Funktion $T$ an den Schluß bringen, wobei es gegebenenfalls vermöge der Funktion $Z$ aus Klammern befreit und, falls es mehrmals vorkommen sollte, vermöge der Funktion $S$ verschmolzen werden muß. So haben wir z. B.:

$$fx \mid^x gxy = fx \mid^x Tgyx = Uf(Tgy).$$

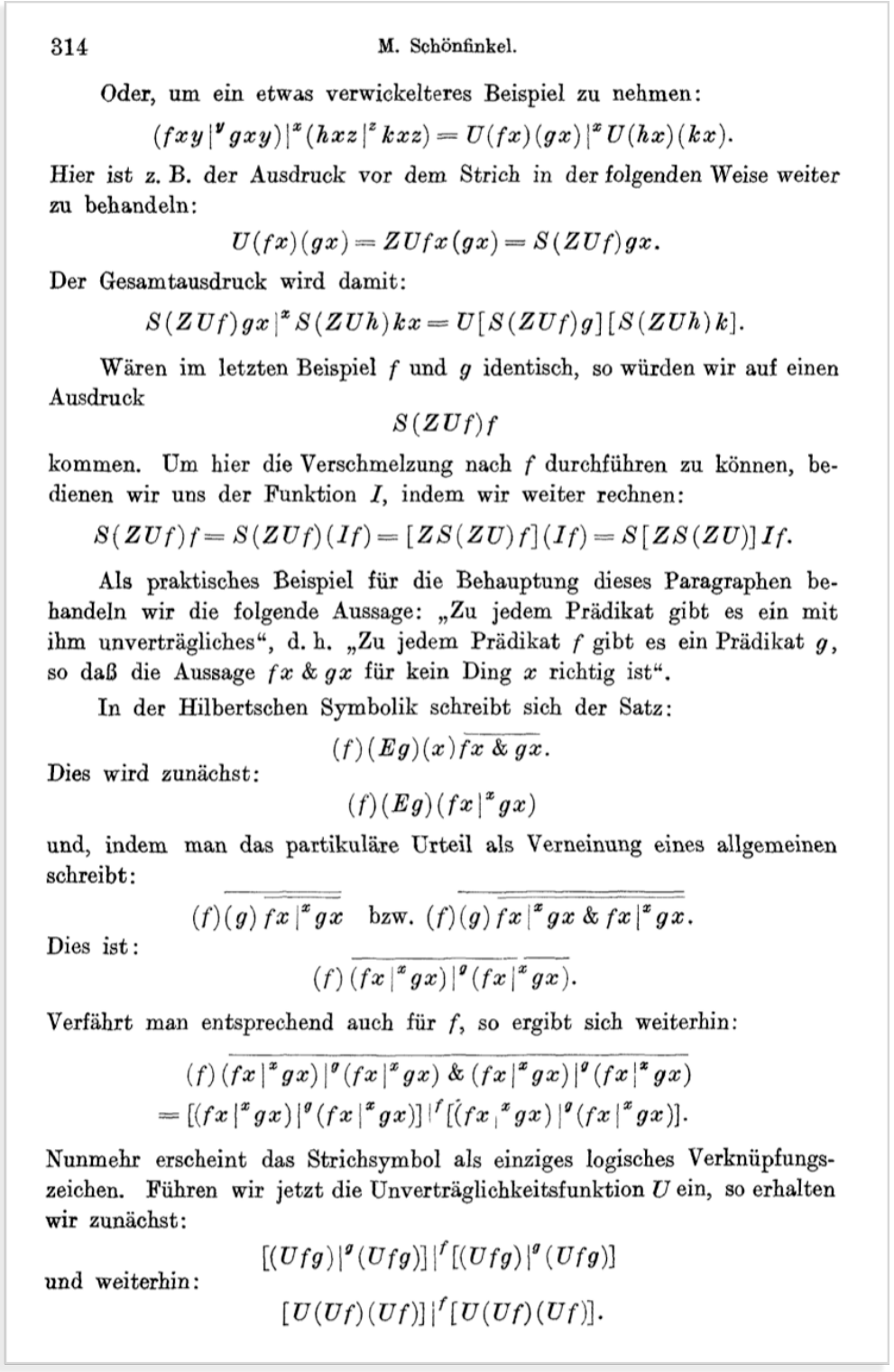
314 M. Schönfinkel.

Oder, um ein etwas verwickelteres Beispiel zu nehmen:

$$(fxy \mid^{y} gxy) \mid^{x} (hxz \mid^{z} kxz) = U(fx)(gx) \mid^{x} U(hx)(kx).$$

Hier ist z. B. der Ausdruck vor dem Strich in der folgenden Weise weiter zu behandeln:

$$U(fx)(gx) = ZUfx(gx) = S(ZUf)gx.$$

Der Gesamtausdruck wird damit:

$$S(ZUf)gx \mid^{x} S(ZUh)kx = U[S(ZUf)g][S(ZUh)k].$$

Wären im letzten Beispiel $f$ und $g$ identisch, so würden wir auf einen Ausdruck

$$S(ZUf)f$$

kommen. Um hier die Verschmelzung nach $f$ durchführen zu können, bedienen wir uns der Funktion $I$, indem wir weiter rechnen:

$$S(ZUf)f = S(ZUf)(If) = [ZS(ZU)f](If) = S[ZS(ZU)]If.$$

Als praktisches Beispiel für die Behauptung dieses Paragraphen behandeln wir die folgende Aussage: „Zu jedem Prädikat gibt es ein mit ihm unverträgliches", d. h. „Zu jedem Prädikat $f$ gibt es ein Prädikat $g$, so daß die Aussage $fx \,\&\, gx$ für kein Ding $x$ richtig ist".

In der Hilbertschen Symbolik schreibt sich der Satz:

$$(f)(Eg)(x)\overline{fx \,\&\, gx}.$$

Dies wird zunächst:

$$(f)(Eg)(fx \mid^{x} gx)$$

und, indem man das partikuläre Urteil als Verneinung eines allgemeinen schreibt:

$$(f)\overline{(g)\,\overline{fx \mid^{x} gx}} \quad \text{bzw.} \quad (f)\overline{(g)\,\overline{fx \mid^{x} gx \,\&\, fx \mid^{x} gx}}.$$

Dies ist:

$$(f)\,\overline{(fx \mid^{x} gx) \mid^{g} (fx \mid^{x} gx)}.$$

Verfährt man entsprechend auch für $f$, so ergibt sich weiterhin:

$$(f)\,\overline{(fx \mid^{x} gx) \mid^{g} (fx \mid^{x} gx) \,\&\, (fx \mid^{x} gx) \mid^{g} (fx \mid^{x} gx)}$$
$$= [(fx \mid^{x} gx) \mid^{g} (fx \mid^{x} gx)] \mid^{f} [(fx \mid^{x} gx) \mid^{g} (fx \mid^{x} gx)].$$

Nunmehr erscheint das Strichsymbol als einziges logisches Verknüpfungszeichen. Führen wir jetzt die Unverträglichkeitsfunktion $U$ ein, so erhalten wir zunächst:

$$[(Ufg) \mid^{g} (Ufg)] \mid^{f} [(Ufg) \mid^{g} (Ufg)]$$

und weiterhin:

$$[U(Uf)(Uf)] \mid^{f} [U(Uf)(Uf)].$$

"It is a remarkable fact", he says, "that every formula of logic can now be expressed by means... solely of *C*, *S* and *U*." So he's saying that any expression from mathematical logic can be written out as some combinator expression in terms of *S*, *C* (now *K*) and *U*. He says that when there are quantifiers like "for all *x*..." it's always possible to use combinators to get rid of the "bound variables" *x*, etc. He says that he "will not give the complete demonstration here", but rather content himself with an example. (Unfortunately—for reasons of the trajectory of his life that are still quite unclear—he never published his "complete demonstration".)

But, OK, so what had he achieved? He'd basically shown that any expression that might appear in predicate logic (with logical connectives, quantifiers, variables, etc.) could be reduced to an expression purely in terms of the combinators *S*, *C* (now *K*) and *U*.

Did he need the *U*? Not really. But he had to have some way to represent the object with mathematical or logical "meaning" on which his combinators would be acting. Today the obvious thing to do would be to have a representation for true and false. And what's more, to represent these purely in terms of combinators. For example, if we took *K* to represent true, and *SK* (s[k]) to represent false, then And can be represented as *SSK* (s[s][k]), Or as *S*(*SS*) *S*(*SK*) (s[s[s]][s][s[k]]) and Nand as *S*(*S*(*K*(*S*(*SS*(*K*(*KK*))))))*S* (s[s[k[s[s[s][k[k[k]]]]]]][s]). Schönfinkel got amazingly far in reducing everything to his "building blocks". But, yes, he missed this final step.

But given that he'd managed to reduce everything to *S*, *C* and *U* he figured he should try to go further. So he considered an object *J* that would be a single building block of *S* and *C*: *JJ*=*S* and *J*(*JJ*)=*C*.

> Weiter als bis zu den Symbolen *C*, *S* und *U* läßt sich, soviel wir sehen, die Zurückführung nicht ohne Zwang treiben.
>
> Rein schematisch könnte man freilich *C*, *S* und *U* sogar durch eine einzige Funktion ersetzen, indem man die neue Funktion *J* einführte durch die Festsetzung:
>
> $$JC = U, \quad JS = C, \quad Jx = S,$$
>
> wo $x$ jedes von *C* und *S* verschiedene Ding ist. Wir stellen zunächst fest, daß *J* seinerseits von *C* und *S* verschieden ist, da nämlich *J* nur drei, *C* ebenso wie *S* dagegen unendlich viele Funktionswerte annimmt. Wir haben infolgedessen:
>
> $$JJ = S, \quad J(JJ) = JS = C, \quad J[J(JJ)] = JC = U,$$
>
> womit die Zurückführung in der Tat geleistet ist. Doch hat diese wegen ihrer augenscheinlichen Willkür wohl kaum sachliche Bedeutung.

With *S* and *K* one can just point to any piece of an expression and see if it reduces. With *J* it's a bit more complicated. In modern Wolfram Language terms one can state the rules as {j[j][x_][y_][z_] → x[z][y[z]], j[j[j]][x_][y_] → x} (where order matters) but to apply these requires pattern matching "clusters of *J*'s" rather than just looking at single *S*'s and *K*'s at a time.

But even though—as Schönfinkel observed—this "final reduction" to *J* didn't work out, getting everything down to *S* and *K* was already amazing. At the beginning of the paper, Schönfinkel had described his objectives. And then he says "It seems to me remarkable in the extreme that the goal we have just set can be realized also; as it happens, it can be done by a reduction to three fundamental signs." (The paper does say three fundamental signs, presumably counting *U* as well as *S* and *K*.)

I'm sure Schönfinkel expected that to reproduce all the richness of mathematical logic he'd need quite an elaborate set of building blocks. And certainly people like Frege, Whitehead and Russell had used what were eventually very complicated setups. Schönfinkel managed to cut through all the complexity to show that simple building blocks were all that was needed. But then he found something else: that actually just two building blocks (*S* and *K*) were enough.

In modern terms, we'd say that Schönfinkel managed to construct a system capable of universal computation. And that's amazing in itself. But even more amazing is that he found he could do it with such a simple setup.

I'm sure Schönfinkel was extremely surprised. And here I personally feel a certain commonality with him. Because in my own explorations of the computational universe, what I've found over and over again is that it takes only remarkably simple systems to be capable of highly complex behavior—and of universal computation. And even after exploring the computational universe for four decades, I'm still continually surprised at just how simple the systems can be.

For me, this has turned into a general principle—the Principle of Computational Equivalence—and a whole conceptual framework around it. Schönfinkel didn't have anything like that to think in terms of. But he was in a sense a good enough scientist that he still managed to discover what he discovered—that many decades later we can see fits in as another piece of evidence for the Principle of Computational Equivalence.

Looking at Schönfinkel's paper a century later, it's remarkable not only for what it discovers, but also for the clarity and simplicity with which it is presented. A little of the notation is now dated (and of course the original paper is written in German, which is no longer the kind of leading language of scholarship it once was). But for the most part, the paper still seems perfectly modern. Except, of course, that now it could be couched in terms of sym-bolic expressions and computation, rather than mathematical logic.

## What Is Their Mathematics?

Combinators are hard to understand, and it's not clear how many people understood them when they were first introduced—let alone understood their implications. It's not a good sign that when Schönfinkel's paper appeared in 1924 the person who helped prepare it for final publication (Heinrich Behmann) added his own three paragraphs at the end, that were quite confused. And Schönfinkel's sole other published paper—coauthored with Paul Bernays in 1927—didn't even mention combinators, even though they could have very profitably been used to discuss the subject at hand (decision problems in mathematical logic).

But in 1927 combinators (if not perhaps Schönfinkel's recognition for them) had a remarkable piece of good fortune. Schönfinkel's paper was discovered by a certain Haskell Curry—who would then devote more than 50 years to studying what he named "combinators", and to spreading the word about them.

At some level I think one can view the main thrust of what Curry and his disciples did with combinators as an effort to "mathematicize" them. Schönfinkel had presented combinators in a rather straightforward "structural" way. But what was the mathematical interpretation of what he did, and of how combinators work in general? What mathematical formalism could capture Schönfinkel's structural idea of substitution? Just what, for example, was the true notion of equality for combinators?

In the end, combinators are fundamentally computational constructs, full of all the phenom-ena of "unbridled computation"—like undecidability and computational irreducibility. And it's inevitable that mathematics as normally conceived can only go so far in "cracking" them.

But back in the 1920s and 1930s the concept and power of computation was not yet under-stood, and it was assumed that the ideas and tools of mathematics would be the ones to use in analyzing a formal system like combinators. And it wasn't that mathematical methods got absolutely nowhere with combinators.

Unlike cellular automata, or even Turing machines, there's a certain immediate structural complexity to combinators, with their elaborate tree structures, equivalences and so on. And so there was progress to be made—and years of work to be done—in untangling this, without having to face the raw features of full-scale computation, like computational irreducibility.

In the end, combinators are full of computational irreducibility. But they also have layers of computational reducibility, some of which are aligned with the kinds of things mathematics and mathematical logic have been set up to handle. And in this there's a curious resonance with our recent Physics Project.

In our models based on hypergraph rewriting there's also a kind of bedrock of computational irreducibility. But as with combinators, there's a certain immediate structural complex ity to what our models do. And there are layers of computational reducibility associated with this. But the remarkable thing with our models is that some of those layers—and the formalisms one can build to understand them—have an immediate interpretation: they are basically the core theories of twentieth-century physics, namely general relativity and quantum mechanics.

Combinators work sufficiently differently that they don't immediately align with that kind of interpretation. But it's still true that one of the important properties discovered in combinators (namely confluence, related to our idea of causal invariance) turns out to be crucial to our models, their correspondence with physics, and in the end our whole ability to perceive regularity in the universe, even in the face of computational irreducibility.

But let's get back to the story of combinators as it played out after Schönfinkel's paper. Schönfinkel had basically set things up in a novel, very direct, structural way. But Curry wanted to connect with more traditional ideas in mathematical logic, and mathematics in general. And after a first paper (published in 1929) which pretty much just recorded his first thoughts, and his efforts to understand what Schönfinkel had done, Curry was by 1930 starting to do things like formulate axioms for combinators, and hoping to prove general theorems about mathematical properties like equality.

Without the understanding of universal computation and their relationship to it, it wasn't clear yet how complicated it might ultimately be to deal with combinators. And Curry pushed forward, publishing more papers and trying to do things like define set theory using his axioms for combinators. But in 1934 disaster struck. It wasn't something about computation or undecidability; instead it was that Stephen Kleene and J. Barkley Rosser showed the axioms Curry had come up to try and "tighten up Schönfinkel" with were just plain inconsistent.

To Kleene and Rosser it provided more evidence of the need for Russell's (originally quite hacky) idea of types—and led them to more complicated axiom systems, and away from combinators. But Curry was undeterred. He revised his axiom system and continued— ultimately for many decades—to see what could be proved about combinators and things like them using mathematical methods.

But already at the beginning of the 1930s there were bigger things afoot around mathematical logic—which would soon intersect with combinators.

## Gödel's Theorem and Computability

How should one represent the fundamental constructs of mathematics? Back in the 1920s nobody thought seriously about using combinators. And instead there were basically three "big brands": *Principia Mathematica*, set theory and Hilbert's program. Relations were being found, details were being filled in, and issues were being found. But there was a general sense that progress was being made.

Quite where the boundaries might lie wasn't clear. For example, could one specify a way to "construct any function" from lower-level primitives? The basic idea of recursion was very old (think: Fibonacci). But by the early 1920s there was a fairly well-formalized notion of "primitive recursion" in which functions always found their values from earlier values. But could all "mathematical" functions be constructed this way?

By 1926 it was known that this wouldn't work: the Ackermann function was a reasonable "mathematical" function, but it wasn't primitive recursive. It meant that definitions had to be generalized (e.g. to "general recursive functions" that didn't just look back at earlier values, but could "look forward until..." as well). But there didn't seem to be any fundamental problem with the idea that mathematics could just "mechanistically" be built out forever from appropriate primitives.

But in 1931 came Gödel's theorem. There'd been a long tradition of identifying paradoxes and inconsistencies, and finding ways to patch them by changing axioms. But Gödel's theorem was based on Peano's by-then-standard axioms for arithmetic (branded by Gödel as a fragment of *Principia Mathematica*). And it showed there was a fundamental problem.

In essence, Gödel took the paradoxical statement "this statement is unprovable" and showed that it could be expressed purely as a statement of arithmetic—roughly a statement about the existence of solutions to appropriate integer equations. And basically what Gödel had to do to achieve this was to create a "compiler" capable of compiling things like "this statement is unprovable" into arithmetic.

In his paper one can basically see him building up different capabilities (e.g. representing arbitrary expressions as numbers through Gödel numbering, checking conditions using general recursion, etc.)—eventually getting to a "high enough level" to represent the statement he wanted:

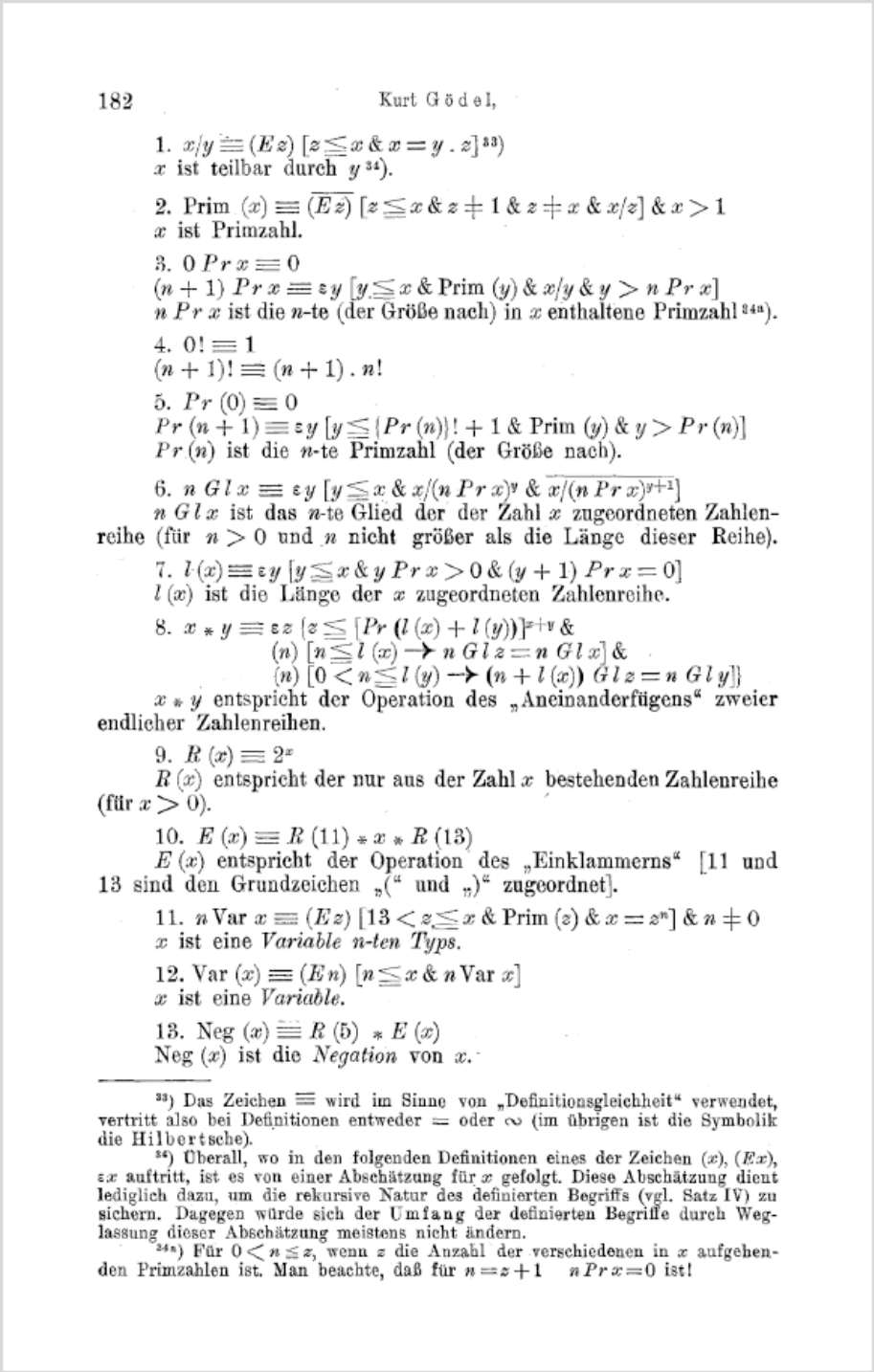
182 Kurt Gödel,

1. $x/y \equiv (Ez)\,[z \leq x \,\&\, x = y \,.\, z]$[33]
$x$ ist teilbar durch $y$[34].

2. $\mathrm{Prim}\,(x) \equiv \overline{(Ez)}\,[z \leq x \,\&\, z \neq 1 \,\&\, z \neq x \,\&\, x/z] \,\&\, x > 1$
$x$ ist Primzahl.

3. $0\, Pr\, x \equiv 0$
$(n+1)\, Pr\, x \equiv \varepsilon y\,[y \leq x \,\&\, \mathrm{Prim}\,(y) \,\&\, x/y \,\&\, y > n\, Pr\, x]$
$n\, Pr\, x$ ist die $n$-te (der Größe nach) in $x$ enthaltene Primzahl[34a].

4. $0! \equiv 1$
$(n+1)! \equiv (n+1)\,.\,n!$

5. $Pr\,(0) \equiv 0$
$Pr\,(n+1) \equiv \varepsilon y\,[y \leq \{Pr\,(n)\}! + 1 \,\&\, \mathrm{Prim}\,(y) \,\&\, y > Pr\,(n)]$
$Pr\,(n)$ ist die $n$-te Primzahl (der Größe nach).

6. $n\, Gl\, x \equiv \varepsilon y\,[y \leq x \,\&\, x/(n\, Pr\, x)^y \,\&\, \overline{x/(n\, Pr\, x)^{y+1}}]$
$n\, Gl\, x$ ist das $n$-te Glied der der Zahl $x$ zugeordneten Zahlenreihe (für $n > 0$ und $n$ nicht größer als die Länge dieser Reihe).

7. $l\,(x) \equiv \varepsilon y\,[y \leq x \,\&\, y\, Pr\, x > 0 \,\&\, (y+1)\, Pr\, x = 0]$
$l\,(x)$ ist die Länge der $x$ zugeordneten Zahlenreihe.

8. $x * y \equiv \varepsilon z\,\{z \leq [Pr\,(l\,(x) + l\,(y))]^{x+y} \,\&$
$(n)\,[n \leq l\,(x) \rightarrow n\, Gl\, z = n\, Gl\, x] \,\&$
$(n)\,[0 < n \leq l\,(y) \rightarrow (n + l\,(x))\, Gl\, z = n\, Gl\, y]\}$
$x * y$ entspricht der Operation des „Aneinanderfügens" zweier endlicher Zahlenreihen.

9. $R\,(x) \equiv 2^x$
$R\,(x)$ entspricht der nur aus der Zahl $x$ bestehenden Zahlenreihe (für $x > 0$).

10. $E\,(x) \equiv R\,(11) * x * R\,(13)$
$E\,(x)$ entspricht der Operation des „Einklammerns" [11 und 13 sind den Grundzeichen „(" und „)" zugeordnet].

11. $n\,\mathrm{Var}\, x \equiv (Ez)\,[13 < z \leq x \,\&\, \mathrm{Prim}\,(z) \,\&\, x = z^n] \,\&\, n \neq 0$
$x$ ist eine *Variable $n$-ten Typs.*

12. $\mathrm{Var}\,(x) \equiv (En)\,[n \leq x \,\&\, n\,\mathrm{Var}\, x]$
$x$ ist eine *Variable.*

13. $\mathrm{Neg}\,(x) \equiv R\,(5) * E\,(x)$
$\mathrm{Neg}\,(x)$ ist die *Negation* von $x$.

[33]) Das Zeichen $\equiv$ wird im Sinne von „Definitionsgleichheit" verwendet, vertritt also bei Definitionen entweder $=$ oder $\sim$ (im übrigen ist die Symbolik die Hilbertsche).

[34]) Überall, wo in den folgenden Definitionen eines der Zeichen $(x)$, $(Ex)$, $\varepsilon x$ auftritt, ist es von einer Abschätzung für $x$ gefolgt. Diese Abschätzung dient lediglich dazu, um die rekursive Natur des definierten Begriffs (vgl. Satz IV) zu sichern. Dagegen würde sich der Umfang der definierten Begriffe durch Weglassung dieser Abschätzung meistens nicht ändern.

[34a]) Für $0 < n \leq z$, wenn $z$ die Anzahl der verschiedenen in $x$ aufgehenden Primzahlen ist. Man beachte, daß für $n = z+1$ $\;n\, Pr\, x = 0$ ist!

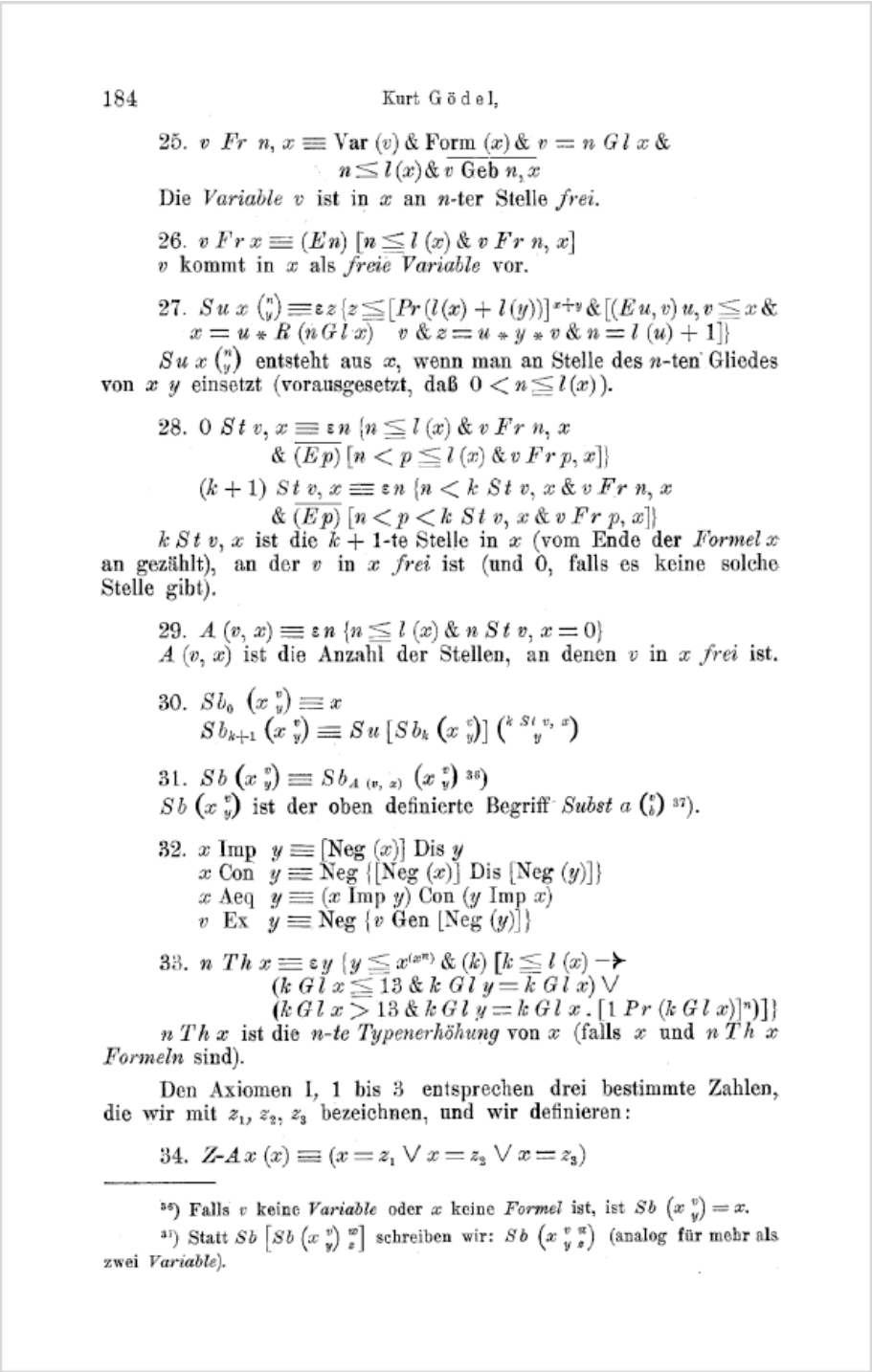
184 Kurt Gödel,

25. $v\; Fr\; n, x \equiv \mathrm{Var}\,(v) \,\&\, \mathrm{Form}\,(x) \,\&\, v = n\, Gl\, x \,\&$
$n \leq l(x) \,\&\, \overline{v\; \mathrm{Geb}\; n, x}$
Die *Variable* $v$ ist in $x$ an $n$-ter Stelle *frei.*

26. $v\, Fr\, x \equiv (En)\,[n \leq l\,(x) \,\&\, v\, Fr\, n, x]$
$v$ kommt in $x$ als *freie Variable* vor.

27. $Su\, x\binom{n}{y} \equiv \varepsilon z\,\{z \leq [Pr\,(l(x) + l\,(y))]^{x+y} \,\&\, [(Eu, v)\, u, v \leq x \,\&$
$x = u * R\,(n\, Gl\, x) * v \,\&\, z = u * y * v \,\&\, n = l\,(u) + 1]\}$
$Su\, x\binom{n}{y}$ entsteht aus $x$, wenn man an Stelle des $n$-ten Gliedes von $x$ $y$ einsetzt (vorausgesetzt, daß $0 < n \leq l(x)$).

28. $0\; St\; v, x \equiv \varepsilon n\,\{n \leq l\,(x) \,\&\, v\, Fr\, n, x$
$\&\, \overline{(Ep)}\,[n < p \leq l\,(x) \,\&\, v\, Fr\, p, x]\}$
$(k+1)\; St\; v, x \equiv \varepsilon n\,\{n < k\; St\; v, x \,\&\, v\, Fr\, n, x$
$\&\, \overline{(Ep)}\,[n < p < k\; St\; v, x \,\&\, v\, Fr\, p, x]\}$
$k\; St\; v, x$ ist die $k+1$-te Stelle in $x$ (vom Ende der *Formel* $x$ an gezählt), an der $v$ in $x$ *frei* ist (und 0, falls es keine solche Stelle gibt).

29. $A\,(v, x) \equiv \varepsilon n\,\{n \leq l\,(x) \,\&\, n\; St\; v, x = 0\}$
$A\,(v, x)$ ist die Anzahl der Stellen, an denen $v$ in $x$ *frei* ist.

30. $Sb_0\left(x\,{v \atop y}\right) \equiv x$
$Sb_{k+1}\left(x\,{v \atop y}\right) \equiv Su\,[Sb_k\left(x\,{v \atop y}\right)]\binom{k\; St\; v, x}{y}$

31. $Sb\left(x\,{v \atop y}\right) \equiv Sb_{A\,(v,\,x)}\left(x\,{v \atop y}\right)$[36]
$Sb\left(x\,{v \atop y}\right)$ ist der oben definierte Begriff *Subst* $a\binom{v}{b}$[37].

32. $x\;\mathrm{Imp}\;y \equiv [\mathrm{Neg}\,(x)]\;\mathrm{Dis}\;y$
$x\;\mathrm{Con}\;y \equiv \mathrm{Neg}\,\{[\mathrm{Neg}\,(x)]\;\mathrm{Dis}\;[\mathrm{Neg}\,(y)]\}$
$x\;\mathrm{Aeq}\;y \equiv (x\;\mathrm{Imp}\;y)\;\mathrm{Con}\;(y\;\mathrm{Imp}\;x)$
$v\;\mathrm{Ex}\;y \equiv \mathrm{Neg}\,\{v\;\mathrm{Gen}\;[\mathrm{Neg}\,(y)]\}$

33. $n\; Th\; x \equiv \varepsilon y\,\{y \leq x^{(x^n)} \,\&\, (k)\,[k \leq l\,(x) \rightarrow$
$(k\, Gl\, x \leq 13 \,\&\, k\, Gl\, y = k\, Gl\, x) \vee$
$(k\, Gl\, x > 13 \,\&\, k\, Gl\, y = k\, Gl\, x\,.\,[1\, Pr\,(k\, Gl\, x)]^n)]\}$
$n\; Th\; x$ ist die *$n$-te Typenerhöhung* von $x$ (falls $x$ und $n\; Th\; x$ *Formeln* sind).

Den Axiomen I, 1 bis 3 entsprechen drei bestimmte Zahlen, die wir mit $z_1, z_2, z_3$ bezeichnen, und wir definieren:

34. $Z\text{-}Ax\,(x) \equiv (x = z_1 \vee x = z_2 \vee x = z_3)$

[36]) Falls $v$ keine *Variable* oder $x$ keine *Formel* ist, ist $Sb\left(x\,{v \atop y}\right) = x$.

[37]) Statt $Sb\left[Sb\left(x\,{v \atop y}\right){w \atop z}\right]$ schreiben wir: $Sb\left(x\,{v \atop y}\,{w \atop z}\right)$ (analog für mehr als zwei *Variable*).

What did Gödel's theorem mean? For the foundations of mathematics it meant that the idea of mechanically proving "all true theorems of mathematics" wasn't going to work. Because it showed that there was at least one statement that by its own admission couldn't be proved, but was still a "statement about arithmetic", in the sense that it could be "compiled into arithmetic".

That was a big deal for the foundations of mathematics. But actually there was something much more significant about Gödel's theorem, even though it wasn't recognized at the time. Gödel had used the primitives of number theory and logic to build what amounted to a computational system—in which one could take things like "this statement is unprovable", and "run them in arithmetic".

What Gödel had, though, wasn't exactly a streamlined general system (after all, it only really needed to handle one statement). But the immediate question then was: if there's a problem with this statement in arithmetic, what about Hilbert's general "decision problem" (*Entscheidungsproblem*) for any axiom system?

To discuss the "general decision problem", though, one needed some kind of general notion of how one could decide things. What ultimate primitives should one use? Schönfinkel (with Paul Bernays)—in his sole other published paper—wrote about a restricted case of the decision problem in 1927, but doesn't seem to have had the idea of using combinators to study it.

By 1934 Gödel was talking about general recursiveness (i.e. definability through general recursion). And Alonzo Church and Stephen Kleene were introducing λ definability. Then in 1936 Alan Turing introduced Turing machines. All these approaches involved setting up certain primitives, then showing that a large class of things could be "compiled" to those primitives. And that—in effect by thinking about having it compile itself—Hilbert's *Entscheidungsproblem* couldn't be solved.

Perhaps no single result along these lines would have been so significant. But it was soon established that all three kinds of systems were exactly equivalent: the set of computations they could represent were the same, as established by showing that one system could emulate another. And from that discovery eventually emerged the modern notion of universal computation—and all its implications for technology and science.

In the early days, though, there was actually a fourth equivalent kind of system—based on string rewriting—that had been invented by Emil Post in 1920–1. Oh, and then there were combinators.

## Lambda Calculus

What was the right "language" to use for setting up mathematical logic? There'd been gradual improvement since the complexities of *Principia Mathematica*. But around 1930 Alonzo Church wanted a new and cleaner setup. And he needed to have a way (as Frege and *Principia Mathematica* had done before him) to represent "pure functions". And that's how he came to invent λ.

Today in the Wolfram Language we have **Function**[x, f[x]] or x⟼f[x] (or various shorthands). Church originally had $\lambda x[\mathbf{M}]$:

An occurrence of a variable **x** in a given formula is called an occurrence of **x** as a *bound variable* in the given formula if it is an occurrence of **x** in a part of the formula of the form λ**x**[**M**]; that is, if there is a formula **M** such that λ**x**[**M**] occurs in the given formula and the occurrence of **x** in question is an occurrence in λ**x**[**M**]. All other occurrences of a variable in a formula are called occurrences as a *free variable.*

A formula is said to be *well-formed* if it is a variable, or if it is one of the symbols *Π*, *Σ*, &, ∼, ι, *Δ*, or if it is obtainable from these symbols by repeated combinations of them of one of the forms {**M**}(**N**) and λ**x**[**M**], where **x** is any variable and **M** and **N** are symbols or formulas which are being combined. This is a definition by induction. It implies the following rules: (1) a variable is well-formed (2) *Π*, *Σ*, &, ∼, ι, and *Δ* are well-formed (3) if **M** and **N** are well-formed then {**M**}(**N**) is well-formed (4) if **x** is a variable and **M** is well-formed then λ**x**[**M**] is well-formed.

All the formulas which will be provable as consequences of our postulates will be well-formed and will contain no free variables.

The undefined terms of a formal system have, as we have explained, no meaning except in connection with a particular application of the system. But for the formal system which we are engaged in constructing we have in mind a particular application, which constitutes, in fact, the motive for constructing it, and we give here the meanings which our undefined terms are to have in this intended application.

If **F** is a function and **A** is a value of the independent variable for which the function is defined, then {**F**}(**A**) represents the value taken on by the function **F** when the independent variable takes on the value **A**. The usual notation is **F**(**A**). We introduce the braces on account of the possibility that **F** might be a combination of several symbols, but, in the case that **F** is a single symbol, we shall often use the notation **F**(**A**) as an abbreviation for the fuller expression.[10]

Adopting a device due to Schönfinkel,[11] we treat a function of two variables as a function of one variable whose values are functions of one variable, and a function of three or more variables similarly. Thus, what is usually written **F**(**A**,**B**) we write {{**F**}(**A**)}(**B**), and what is usually written **F**(**A**, **B**, **C**) we write {{{**F**}(**A**)}(**B**)}(**C**), and so on. But again we frequently find it convenient to employ the more usual notations as abbreviations.

If **M** is any formula containing the variable **x**, then λ**x**[**M**] is a symbol for the function whose values are those given by the formula. That is,

[10] The braces { } are, as a matter of fact, superfluous and might have been omitted from our list of undefined terms, but their inclusion makes for readability of formulas.

[11] M. Schönfinkel, Über die Bausteine der mathematischen Logik, Math. Annalen, vol. 92 (1924), pp. 305–316.

But what's perhaps most notable is that on the very first page he defines λ, he's referencing Schönfinkel's combinator paper. (Well, specifically, he's referencing it because he wants to use the device Schönfinkel invented that we now call currying—f[x][y] in place of f[x,y]—though ironically he doesn't mention Curry.) In his 1932 paper (apparently based on work in 1928–9) λ is almost a sideshow—the main event being the introduction of 37 formal postulates for mathematical logic:

356 A. CHURCH.

III. *If* **I** *is true, if* **M** *and* **N** *are well-formed, if the variable* **x** *occurs in* **M**, *and if the bound variables in* **M** *are distinct both from* **x** *and from the free variables in* **N**, *then* **K**, *the result of substituting* $\{\lambda \mathbf{x} . \mathbf{M}\}(\mathbf{N})$ *for a particular occurrence of* $S^{\mathbf{x}}_{\mathbf{N}}\mathbf{M}|$ *in* **I**, *is also true.*

IV. *If* $\{\mathbf{F}\}(\mathbf{A})$ *is true and* **F** *and* **A** *are well-formed, then* $\Sigma(\mathbf{F})$ *is true.*

V. *If* $\Pi(\mathbf{F}, \mathbf{G})$ *and* $\{\mathbf{F}\}(\mathbf{A})$ *are true, and* **F**, **G**, *and* **A** *are well-formed, then* $\{\mathbf{G}\}(\mathbf{A})$ *is true.*

And our formal postulates are the thirty-seven following:

1. $\Sigma(\varphi) \supset_\varphi \Pi(\varphi, \varphi)$.
2. $'x . \varphi(x) \supset_\varphi . \Pi(\varphi, \psi) \supset_\psi \psi(x)$.
3. $\Sigma(\sigma) \supset_\sigma . [\sigma(x) \supset_x \varphi(x)] \supset_\varphi . \Pi(\varphi, \psi) \supset_\psi . \sigma(x) \supset_x \psi(x)$.
4. $\Sigma(\varrho) \supset_\varrho . \Sigma y [\varrho(x) \supset_x \varphi(x, y)] \supset_\varphi . [\varrho(x) \supset_x \Pi(\varphi(x), \psi(x))] \supset_\psi .$ $[\varrho(x) \supset_x \varphi(x, y)] \supset_y . \varrho(x) \supset_x \psi(x, y)$.
5. $\Sigma(\varphi) \supset_\varphi . \Pi(\varphi, \psi) \supset_\psi . \varphi(f(x)) \supset_{fx} \psi(f(x))$.
6. $'x . \varphi(x) \supset_\varphi . \Pi(\varphi, \psi(x)) \supset_\psi \psi(x, x)$.
7. $\varphi(x, f(x)) \supset_{\varphi fx} . \Pi(\varphi(x), \psi(x)) \supset_\psi \psi(x, f(x))$.
8. $\Sigma(\varrho) \supset_\varrho . \Sigma y [\varrho(x) \supset_x \varphi(x, y)] \supset_\varphi . [\varrho(x) \supset_x \Pi(\varphi(x), \psi)] \supset_\psi .$ $[\varrho(x) \supset_x \varphi(x, y)] \supset_y \psi(y)$.
9. $'x . \varphi(x) \supset_\varphi \Sigma(\varphi)$.
10. $\Sigma x \varphi(f(x)) \supset_{f\varphi} \Sigma(\varphi)$.
11. $\varphi(x, x) \supset_{\varphi x} \Sigma(\varphi(x))$.
12. $\Sigma(\varphi) \supset_\varphi \Sigma x \varphi(x)$.
13. $\Sigma(\varphi) \supset_\varphi . [\varphi(x) \supset_x \psi(x)] \supset_\psi \Pi(\varphi, \psi)$.
14. $p \supset_p . q \supset_q pq$.
15. $pq \supset_{qp} p$.
16. $pq \supset_{pq} q$.
17. $\Sigma x \Sigma \theta [\varphi(x) . \sim \theta(x) . \Pi(\psi, \theta)] \supset_{\varphi\psi} \sim \Pi(\varphi, \psi)$.
18. $\sim \Pi(\varphi, \psi) \supset_{\varphi\psi} \Sigma x \Sigma \theta . \varphi(x) . \sim \theta(x) . \Pi(\psi, \theta)$.
19. $\Sigma x \Sigma \theta [\sim \varphi(u, x) . \sim \theta(u) . \Sigma(\varphi(y)) \supset_y \theta(y)] \supset_{\varphi u} \sim \Sigma(\varphi(u))$.
20. $\sim \Sigma(\varphi) \supset_\varphi \Sigma x . \sim \varphi(x)$.
21. $p \supset_p . \sim q \supset_q \sim . pq$.
22. $\sim p \supset_p . q \supset_q \sim . pq$.
23. $\sim p \supset_p . \sim q \supset_q \sim . pq$.
24. $p \supset_p . [\sim . pq] \supset_q \sim q$.

By the next year J. Barkley Rosser is trying to retool Curry's "combinatory logic" with combinators of his own—and showing how they correspond to lambda expressions:

**We shall use Church's method for denoting definitions (see Church 1932, p. 355) and shall list the following, giving on the right the equivalent in Church's notation:**

| | | |
|---|---|---|
| $T \rightarrow JII$ | $\lambda xf \cdot f(x)$ | (See footnote 4) |
| $C \rightarrow JT(JT)(JT)$ | $\lambda fxy \cdot f(y, x)$ | |
| $B \rightarrow C(JIC)(JI)$ | $\lambda fgx \cdot f(g(x))$ | |
| $W \rightarrow C(C(BC(C(BJT)T))T)$ | $\lambda fx \cdot f(x, x)$ | |
| $1 \rightarrow BI$ | $\lambda fx \cdot f(x)$ | |
| $\mathbf{p} \times \mathbf{q} \rightarrow B\mathbf{pq}$ | $\lambda x \cdot \mathbf{p}(\mathbf{q}(x))$ | |

[4] $\lambda xf \cdot f(x)$ is an abbreviation of $\lambda x \lambda f \cdot f(x)$.

Then in 1935 lambda calculus has its big "coming out" in Church's "An Unsolvable Problem of Elementary Number Theory", in which he introduces the idea that any "effectively calculable" function should be "$\lambda$ definable", then defines integers in terms of $\lambda$'s ("Church numerals")

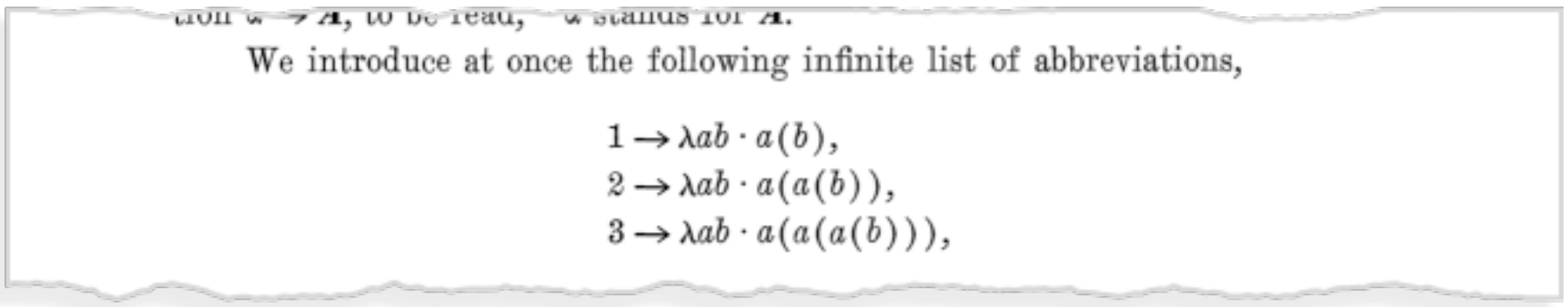
We introduce at once the following infinite list of abbreviations,

$$1 \rightarrow \lambda ab \cdot a(b),$$
$$2 \rightarrow \lambda ab \cdot a(a(b)),$$
$$3 \rightarrow \lambda ab \cdot a(a(a(b))),$$

and then shows that the problem of determining equivalence for $\lambda$ expressions is undecidable.

Very soon thereafter Turing publishes his "On Computable Numbers, with an Application to the *Entscheidungsproblem*" in which he introduces his much more manifestly mechanistic Turing machine model of computation. In the main part of the paper there are no lambdas—or combinators—to be seen. But by late 1936 Turing had gone to Princeton to be a student with Church—and added a note showing the correspondence between his Turing machines and Church's lambda calculus.

By the next year, when Turing is writing his rather abstruse "Systems of Logic Based on Ordinals" he's using lambda calculus all over the place. Early in the document he writes $I \rightarrow \lambda x[x]$, and soon he's mixing lambdas and combinators with wild abandon—and in fact he'd already published a one-page paper which introduced the fixed-point combinator Θ (and, yes, the *K* in the title refers to Schönfinkel's *K* combinator):

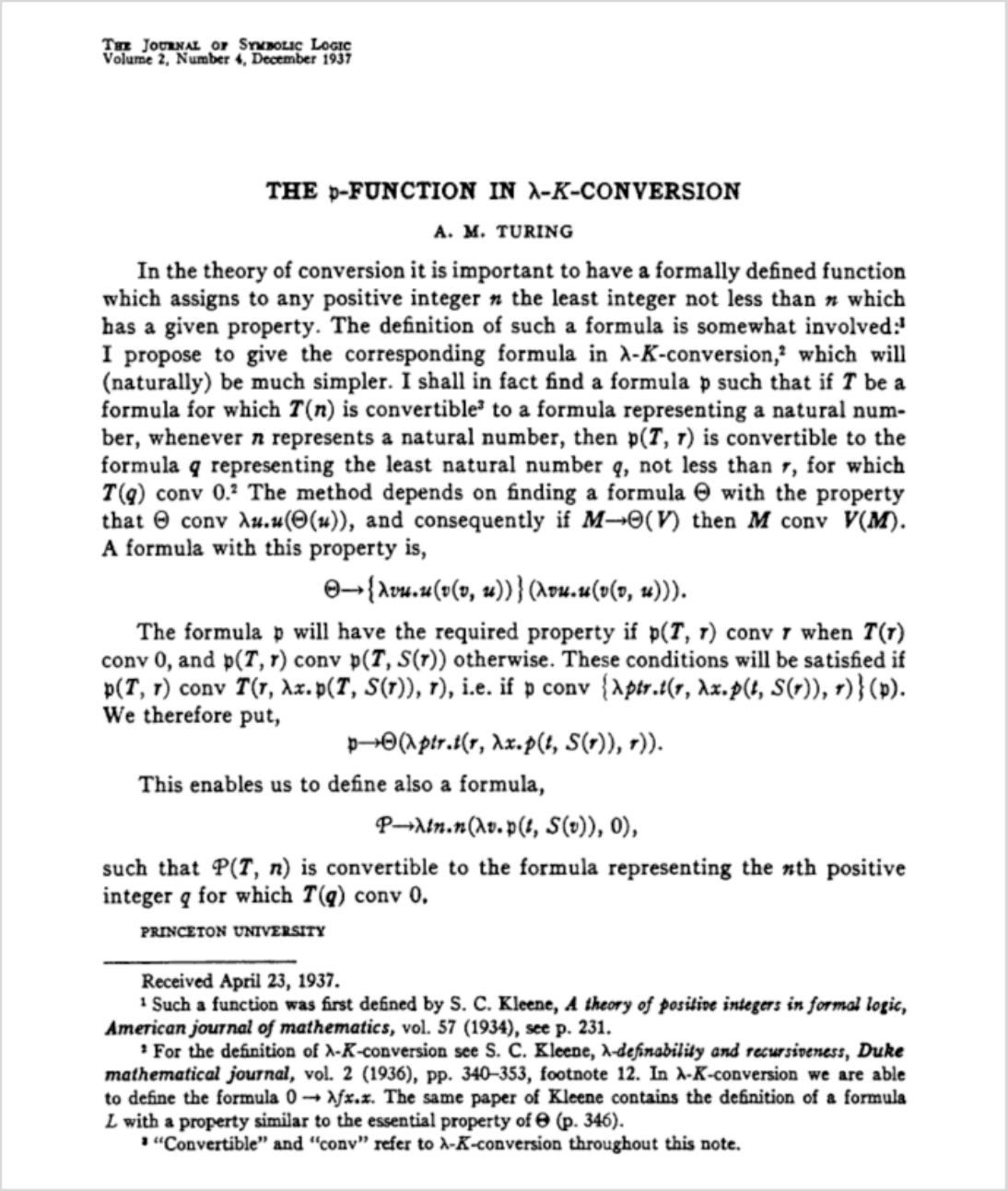

**THE 𝔭-FUNCTION IN λ-*K*-CONVERSION**

A. M. TURING

In the theory of conversion it is important to have a formally defined function which assigns to any positive integer $n$ the least integer not less than $n$ which has a given property. The definition of such a formula is somewhat involved:[1] I propose to give the corresponding formula in λ-*K*-conversion,[2] which will (naturally) be much simpler. I shall in fact find a formula 𝔭 such that if $T$ be a formula for which $T(n)$ is convertible[3] to a formula representing a natural number, whenever $n$ represents a natural number, then $\mathfrak{p}(T, r)$ is convertible to the formula $q$ representing the least natural number $q$, not less than $r$, for which $T(q)$ conv 0.[2] The method depends on finding a formula Θ with the property that Θ conv $\lambda u.u(\Theta(u))$, and consequently if $M \rightarrow \Theta(V)$ then $M$ conv $V(M)$. A formula with this property is,

$$\Theta \rightarrow \{\lambda vu.u(v(v, u))\}(\lambda vu.u(v(v, u))).$$

The formula 𝔭 will have the required property if $\mathfrak{p}(T, r)$ conv $r$ when $T(r)$ conv 0, and $\mathfrak{p}(T, r)$ conv $\mathfrak{p}(T, S(r))$ otherwise. These conditions will be satisfied if $\mathfrak{p}(T, r)$ conv $T(r, \lambda x.\mathfrak{p}(T, S(r)), r)$, i.e. if 𝔭 conv $\{\lambda ptr.t(r, \lambda x.p(t, S(r)), r)\}(\mathfrak{p})$. We therefore put,

$$\mathfrak{p} \rightarrow \Theta(\lambda ptr.t(r, \lambda x.p(t, S(r)), r)).$$

This enables us to define also a formula,

$$\mathcal{P} \rightarrow \lambda tn.n(\lambda v.\mathfrak{p}(t, S(v)), 0),$$

such that $\mathcal{P}(T, n)$ is convertible to the formula representing the $n$th positive integer $q$ for which $T(q)$ conv 0.

PRINCETON UNIVERSITY

[1] Such a function was first defined by S. C. Kleene, *A theory of positive integers in formal logic*, *American journal of mathematics*, vol. 57 (1934), see p. 231.

[2] For the definition of λ-*K*-conversion see S. C. Kleene, *λ-definability and recursiveness*, *Duke mathematical journal*, vol. 2 (1936), pp. 340–353, footnote 12. In λ-*K*-conversion we are able to define the formula $0 \rightarrow \lambda fx.x$. The same paper of Kleene contains the definition of a formula $L$ with a property similar to the essential property of Θ (p. 346).

[3] "Convertible" and "conv" refer to λ-*K*-conversion throughout this note.

When Church summarized the state of lambda calculus in 1941 in his "The Calculi of Lambda-Conversion" he again made extensive use of combinators. Schönfinkel's *K* is prominent. But Schönfinkel's S is nowhere to be seen—and in fact Church has his own *S* combinator $S[n][f][x] \to f[n[f][x]]$ which implements successors in Church's numeral system. And he has also has a few other "basic combinators" that he routinely uses.

In the end, combinators and lambda calculus are completely equivalent, and it's quite easy to convert between them—but there's a curious tradeoff. In lambda calculus one names variables, which is good for human readability, but can lead to problems at a formal level. In combinators, things are formally much cleaner, but the expressions one gets can be completely incomprehensible to humans.

The point is that in a lambda expression like $\lambda x\, \lambda y\, x[y]$ one's naming the variables (here $x$ and $y$), but really these names are just placeholders: what they are doesn't matter; they're just showing where different arguments go. And in a simple case like this, everything is fine. But what happens if one substitutes for y another lambda expression, say $\lambda x\, f[x]$? What is that $x$? Is it the same $x$ as the one outside, or something different? In practice, there are all sorts of renaming schemes that can be used, but they tend to be quite hacky, and things can quickly get tangled up. And if one wants to make formal proofs about lambda calculus, this can potentially be a big problem, and indeed at the beginning it wasn't clear it wouldn't derail the whole idea of lambda calculus.

And that's part of why the correspondence between lambda calculus and combinators was important. With combinators there are no variables, and so no variable names to get tangled up. So if one can show that something can be converted to combinators—even if one never looks at the potentially very long and ugly combinator expression that's generated—one knows one's safe from issues about variable names.

There are still plenty of other complicated issues, though. Prominent among them are questions about when combinator expressions can be considered equal. Let's say you have a combinator expression, like s[s[s[s][k]]][k]. Well, you can repeatedly apply the rules for combinators to transform and reduce it. And it'll often end up at a fixed point, where no rules apply anymore. But a basic question is whether it matters in which order the rules are applied. And in 1936 Church and Rosser proved it doesn't.

Actually, what they specifically proved was the analogous result for lambda calculus. They drew a picture to indicate different possible orders in which lambdas could be reduced out, and showed it didn't matter which path one takes:

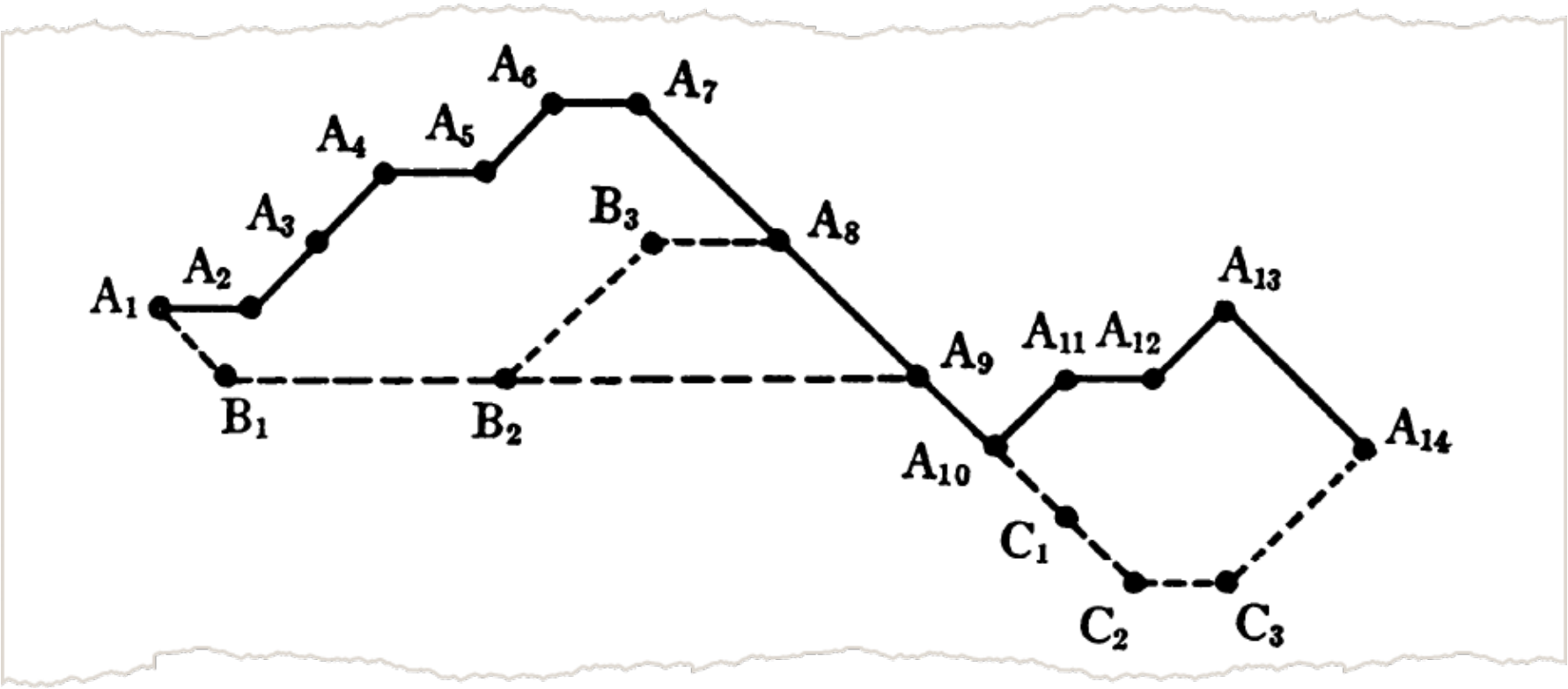

This all might seem like a detail. But it turns out that generalizations of their result apply to all sorts of systems. In doing computations (or automatically proving theorems) it's all about "it doesn't matter what path you take; you'll always get the same result". And that's important. But recently there's been another important application that's shown up. It turns out that a generalization of the "Church–Rosser property" is what we call causal invariance in our Physics Project.

And it's causal invariance that leads in our models to relativistic invariance, general covariance, objective reality in quantum mechanics, and other central features of physics.

## Practical Computation

In retrospect, one of the great achievements of the 1930s was the inception of what ended up being the idea of universal computation. But at the time what was done was couched in terms of mathematical logic and it was far from obvious that any of the theoretical structures being built would have any real application beyond thinking about the foundations of mathematics. But even as people like Hilbert were talking in theoretical terms about the mechanization of mathematics, more and more there were actual machines being built for doing mathematical calculations.

We know that even in antiquity (at least one) simple gear-based mechanical calculational devices existed. In the mid-1600s arithmetic calculators started being constructed, and by the late 1800s they were in widespread use. At first they were mechanical, but by the 1930s most were electromechanical, and there started to be systems where units for carrying out different arithmetic operations could be chained together. And by the end of the 1940s fairly elaborate such systems based on electronics were being built.

Already in the 1830s Charles Babbage had imagined an "analytical engine" which could do different operations depending on a "program" specified by punch cards—and Ada Lovelace had realized that such a machine had broad "computational" potential. But by the 1930s a century had passed and nothing like this was connected to the theoretical developments that were going on—and the actual engineering of computational systems was done without any particular overarching theoretical framework.

Still, as electronic devices got more complicated and scientific interest in psychology intensified, something else happened: there started to be the idea (sometimes associated with the name cybernetics) that somehow electronics might reproduce how things like brains work. In the mid-1930s Claude Shannon had shown that Boolean algebra could represent how switching circuits work, and in 1943 Warren McCulloch and Walter Pitts proposed a model of idealized neural networks formulated in something close to mathematical logic terms.

Meanwhile by the mid-1940s John von Neumann—who had worked extensively on mathematical logic—had started suggesting math-like specifications for practical electronic computers, including the way their programs might be stored electronically. At first he made lots of brain-like references to "organs" and "inhibitory connections", and essentially no mention of ideas from mathematical logic. But by the end of the 1940s von Neumann was talking at least conceptually about connections to Gödel's theorem and Turing machines, Alan Turing had become involved with actual electronic computers, and there was the beginning of widespread understanding of the notion of general-purpose computers and universal computation.

In the 1950s there was an explosion of interest in what would now be called the theory of computation—and great optimism about its relevance to artificial intelligence. There was all sorts of "interdisciplinary work" on fairly "concrete" models of computation, like finite automata, Turing machines, cellular automata and idealized neural networks. More "abstract" approaches, like recursive functions, lambda calculus—and combinators—remained, however, pretty much restricted to researchers in mathematical logic.

When early programming languages started to appear in the latter part of the 1950s, thinking about practical computers began to become a bit more abstract. It was understood that the grammars of languages could be specified recursively—and actual recursion (of functions being able to call themselves) just snuck into the specification of ALGOL 60. But what about the structures on which programs operated? Most of the concentration was on arrays (sometimes rather elegantly, as in APL) and, occasionally, character strings.

But a notable exception was LISP, described in John McCarthy's 1960 paper "Recursive Functions of Symbolic Expressions and Their Computation by Machine, Part I" (part 2 was not written). There was lots of optimism about AI at the time, and the idea was to create a language to "implement AI"—and do things like "mechanical theorem proving". A key idea—that McCarthy described as being based on "recursive function formalism"—was to have tree-structured symbolic expressions ("*S* expressions"). (In the original paper, what's now Wolfram Language–style f[g[x]] "*M* expression" notation, complete with square brackets, was used as part of the specification, but the quintessential-LISP-like (*f* (*g* *x*)) notation won out when LISP was actually implemented.)

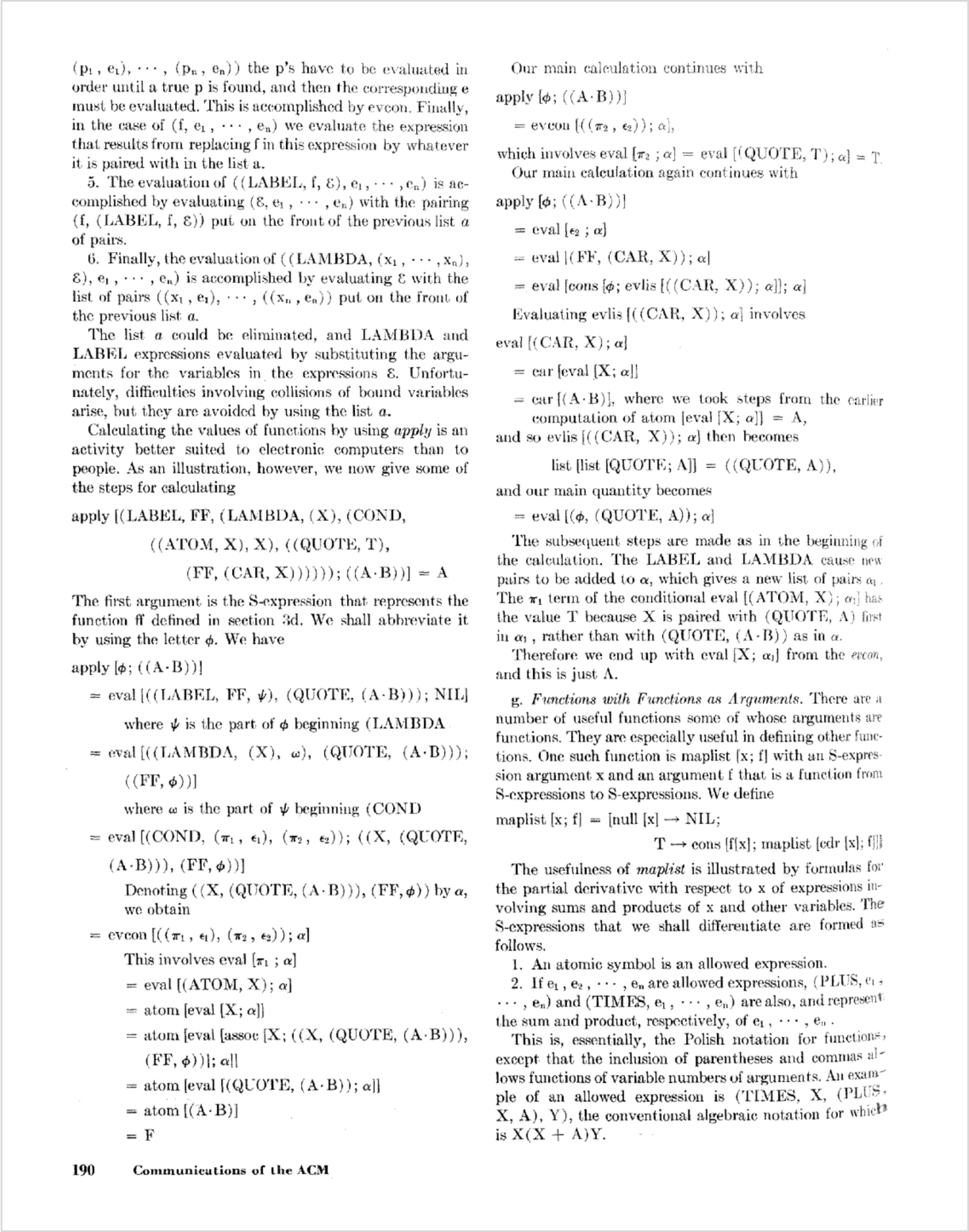

$(p_1, e_1), \cdots, (p_n, e_n))$ the p's have to be evaluated in order until a true p is found, and then the corresponding e must be evaluated. This is accomplished by evcon. Finally, in the case of $(f, e_1, \cdots, e_n)$ we evaluate the expression that results from replacing f in this expression by whatever it is paired with in the list a.

5. The evaluation of $((\text{LABEL}, f, \mathcal{E}), e_1, \cdots, e_n)$ is accomplished by evaluating $(\mathcal{E}, e_1, \cdots, e_n)$ with the pairing $(f, (\text{LABEL}, f, \mathcal{E}))$ put on the front of the previous list *a* of pairs.

6. Finally, the evaluation of $((\text{LAMBDA}, (x_1, \cdots, x_n), \mathcal{E}), e_1, \cdots, e_n)$ is accomplished by evaluating $\mathcal{E}$ with the list of pairs $((x_1, e_1), \cdots, ((x_n, e_n))$ put on the front of the previous list *a*.

The list *a* could be eliminated, and LAMBDA and LABEL expressions evaluated by substituting the arguments for the variables in the expressions $\mathcal{E}$. Unfortunately, difficulties involving collisions of bound variables arise, but they are avoided by using the list *a*.

Calculating the values of functions by using *apply* is an activity better suited to electronic computers than to people. As an illustration, however, we now give some of the steps for calculating apply [(LABEL, FF, (LAMBDA, (X), (COND,
((ATOM, X), X), ((QUOTE, T),
(FF, (CAR, X)))))); ((A·B))] = A

The first argument is the S-expression that represents the function ff defined in section 3d. We shall abbreviate it by using the letter $\phi$. We have

apply [$\phi$; ((A·B))]
= eval [((LABEL, FF, $\psi$), (QUOTE, (A·B))); NIL]
where $\psi$ is the part of $\phi$ beginning (LAMBDA
= eval [((LAMBDA, (X), $\omega$), (QUOTE, (A·B))); ((FF, $\phi$))]
where $\omega$ is the part of $\psi$ beginning (COND
= eval [(COND, ($\pi_1$, $\epsilon_1$), ($\pi_2$, $\epsilon_2$)); ((X, (QUOTE, (A·B))), (FF, $\phi$))]
Denoting ((X, (QUOTE, (A·B))), (FF, $\phi$)) by $\alpha$, we obtain
= evcon [(($\pi_1$, $\epsilon_1$), ($\pi_2$, $\epsilon_2$)); $\alpha$]
This involves eval [$\pi_1$ ; $\alpha$]
= eval [(ATOM, X); $\alpha$]
= atom [eval [X; $\alpha$]]
= atom [eval [assoc [X; ((X, (QUOTE, (A·B))), (FF, $\phi$))]; $\alpha$]]
= atom [eval [(QUOTE, (A·B)); $\alpha$]]
= atom [(A·B)]
= F

Our main calculation continues with

apply [$\phi$; ((A·B))]
= evcon [(($\pi_2$, $\epsilon_2$)); $\alpha$],

which involves eval [$\pi_2$ ; $\alpha$] = eval [(QUOTE, T); $\alpha$] = T.

Our main calculation again continues with

apply [$\phi$; ((A·B))]
= eval [$\epsilon_2$ ; $\alpha$]
= eval [(FF, (CAR, X)); $\alpha$]
= eval [cons [$\phi$; evlis [((CAR, X)); $\alpha$]]; $\alpha$]

Evaluating evlis [((CAR, X)); $\alpha$] involves

eval [(CAR, X); $\alpha$]
= car [eval [X; $\alpha$]]
= car [(A·B)], where we took steps from the earlier computation of atom [eval [X; $\alpha$]] = A,

and so evlis [((CAR, X)); $\alpha$] then becomes list [list [QUOTE; A]] = ((QUOTE, A)),

and our main quantity becomes

= eval [($\phi$, (QUOTE, A)); $\alpha$]

The subsequent steps are made as in the beginning of the calculation. The LABEL and LAMBDA cause new pairs to be added to $\alpha$, which gives a new list of pairs $\alpha_1$. The $\pi_1$ term of the conditional eval [(ATOM, X); $\alpha_1$] has the value T because X is paired with (QUOTE, A) first in $\alpha_1$, rather than with (QUOTE, (A·B)) as in $\alpha$.

Therefore we end up with eval [X; $\alpha_1$] from the *evcon*, and this is just A.

g. *Functions with Functions as Arguments.* There are a number of useful functions some of whose arguments are functions. They are especially useful in defining other functions. One such function is maplist [x; f] with an S-expression argument x and an argument f that is a function from S-expressions to S-expressions. We define maplist [x; f] = [null [x] $\rightarrow$ NIL;
T $\rightarrow$ cons [f[x]; maplist [cdr [x]; f]]]

The usefulness of *maplist* is illustrated by formulas for the partial derivative with respect to x of expressions involving sums and products of x and other variables. The S-expressions that we shall differentiate are formed as follows.

1. An atomic symbol is an allowed expression.
2. If $e_1, e_2, \cdots, e_n$ are allowed expressions, $(\text{PLUS}, e_1, \cdots, e_n)$ and $(\text{TIMES}, e_1, \cdots, e_n)$ are also, and represent the sum and product, respectively, of $e_1, \cdots, e_n$.

This is, essentially, the Polish notation for functions, except that the inclusion of parentheses and commas allows functions of variable numbers of arguments. An example of an allowed expression is (TIMES, X, (PLUS, X, A), Y), the conventional algebraic notation for which is X(X + A)Y.

190 **Communications of the ACM**

An issue in LISP was how to take "expressions" (which were viewed as representing things) and turn them into functions (which do things). And the basic plan was to use Church's idea of $\lambda$ notation. But when it came time to implement this, there was, of course, trouble with name collisions, which ended up getting handled in quite hacky ways. So did McCarthy know about combinators? The answer is yes, as his 1960 paper shows:

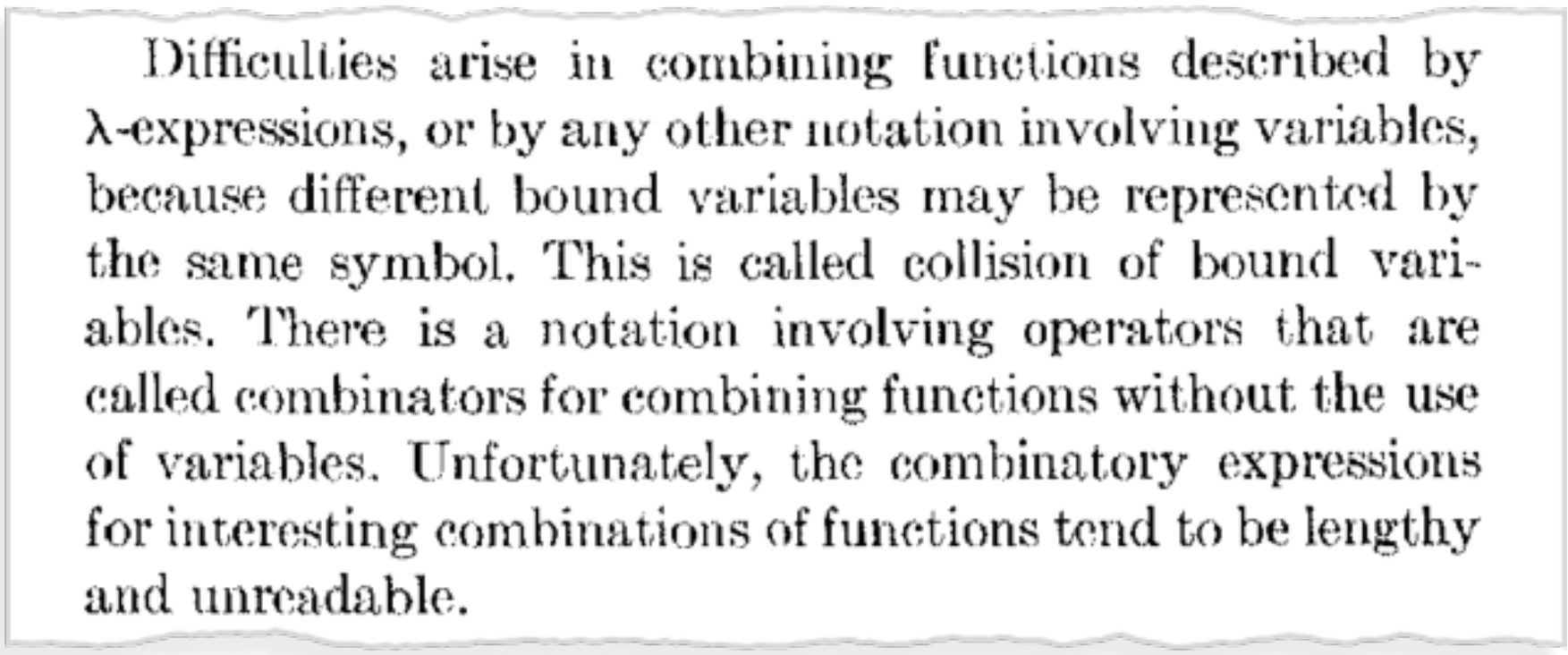
Difficulties arise in combining functions described by λ-expressions, or by any other notation involving variables, because different bound variables may be represented by the same symbol. This is called collision of bound variables. There is a notation involving operators that are called combinators for combining functions without the use of variables. Unfortunately, the combinatory expressions for interesting combinations of functions tend to be lengthy and unreadable.

I actually didn't know until just now that McCarthy had ever even considered combinators, and in the years I knew him I don't think I ever personally talked to him about them. But it seems that for McCarthy—as for Church—combinators were a kind of "comforting backstop" that ensured that it was OK to use lambdas, and that if things went too badly wrong with variable naming, there was at least in principle always a way to untangle everything.

In the practical development of computers and computer languages, even lambdas—let alone combinators—weren't really much heard from again (except in a small AI circle) until the 1980s. And even then it didn't help that in an effort variously to stay close to hardware and to structure programs there tended to be a desire to give everything a "data type"—which was at odds with the "consume any expression" approach of standard combinators and lambdas. But beginning in the 1980s—particularly with the progressive rise of functional programming—lambdas, at least, have steadily gained in visibility and practical application.

What of combinators? Occasionally as a proof of principle there'll be a hardware system developed that natively implements Schönfinkel's combinators. Or—particularly in modern times—there'll be an esoteric language that uses combinators in some kind of purposeful effort at obfuscation. Still, a remarkable cross-section of notable people concerned with the foundations of computing have—at one time or another—taught about combinators or written a paper about them. And in recent years the term "combinator" has become more popular as a way to describe a "purely applicative" function.

But by and large the important ideas that first arose with combinators ended up being absorbed into practical computing by quite circuitous routes, without direct reference to their origins, or to the specific structure of combinators.

## Combinators in Culture

For 100 years combinators have mostly been an obscure academic topic, studied particularly in connection with lambda calculus, at borders between theoretical computer science, mathematical logic and to some extent mathematical formalisms like category theory. Much of the work that's been done can be traced in one way or another to the influence of Haskell Curry or Alonzo Church—particularly through their students, grandstudents, great-grandstudents, etc. Partly in the early years, most of the work was centered in the US, but by the 1960s there was a strong migration to Europe and especially the Netherlands.

But even with all their abstractness and obscurity, on a few rare occasions combinators have broken into something closer to the mainstream. One such time was with the popular logic-puzzle book *To Mock a Mockingbird*, published in 1985 by Raymond Smullyan—a former student of Alonzo Church's. It begins: "A certain enchanted forest is inhabited by talking birds" and goes on to tell a story that's basically about combinators "dressed up" as birds calling each other (*S* is the "starling", *K* the "kestrel")—with a convenient "bird who's who" at the end. The book is dedicated "To the memory of Haskell Curry—an early pioneer in combinatory logic and an avid bird-watcher".

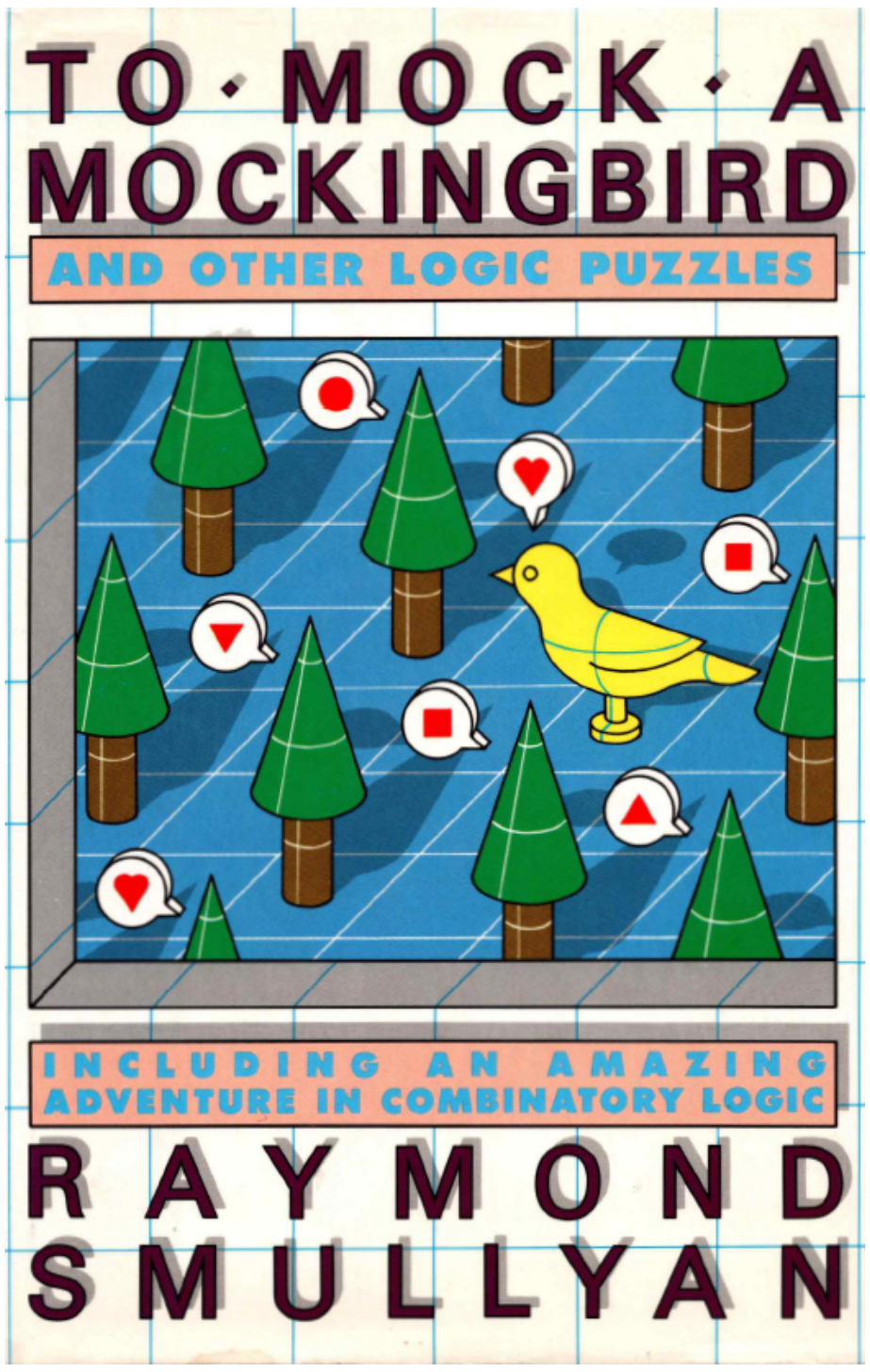

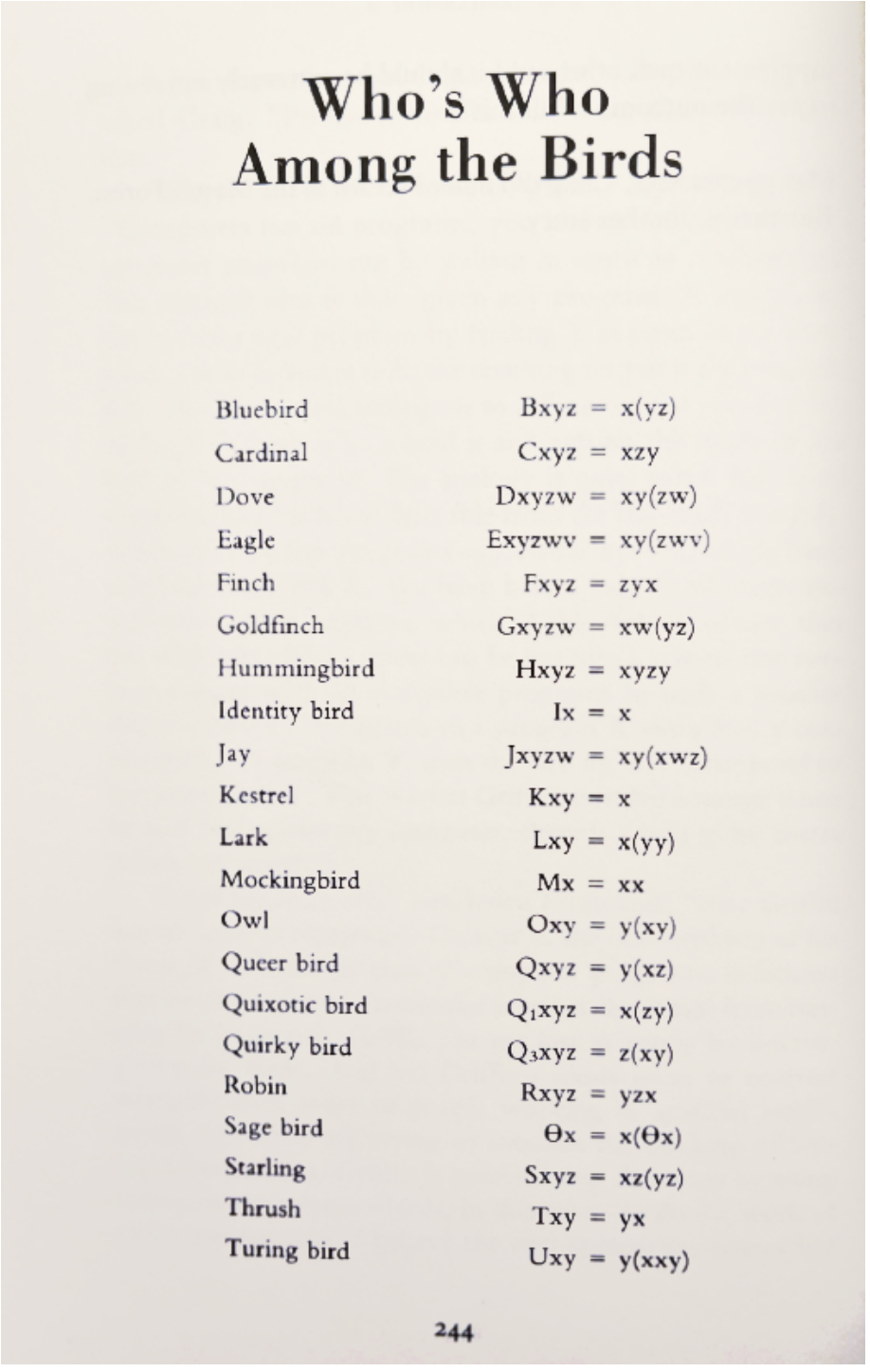

## Who's Who Among the Birds

| | |
|---|---|
| Bluebird | Bxyz = x(yz) |
| Cardinal | Cxyz = xzy |
| Dove | Dxyzw = xy(zw) |
| Eagle | Exyzwv = xy(zwv) |
| Finch | Fxyz = zyx |
| Goldfinch | Gxyzw = xw(yz) |
| Hummingbird | Hxyz = xyzy |
| Identity bird | Ix = x |
| Jay | Jxyzw = xy(xwz) |
| Kestrel | Kxy = x |
| Lark | Lxy = x(yy) |
| Mockingbird | Mx = xx |
| Owl | Oxy = y(xy) |
| Queer bird | Qxyz = y(xz) |
| Quixotic bird | $Q_1$xyz = x(zy) |
| Quirky bird | $Q_3$xyz = z(xy) |
| Robin | Rxyz = yzx |
| Sage bird | Θx = x(Θx) |
| Starling | Sxyz = xz(yz) |
| Thrush | Txy = yx |
| Turing bird | Uxy = y(xxy) |

244

And then there's Y Combinator. The original *Y* combinator arose out of work that Curry did in the 1930s on the consistency of axiom systems for combinators, and it appeared explicitly in his 1958 classic book:

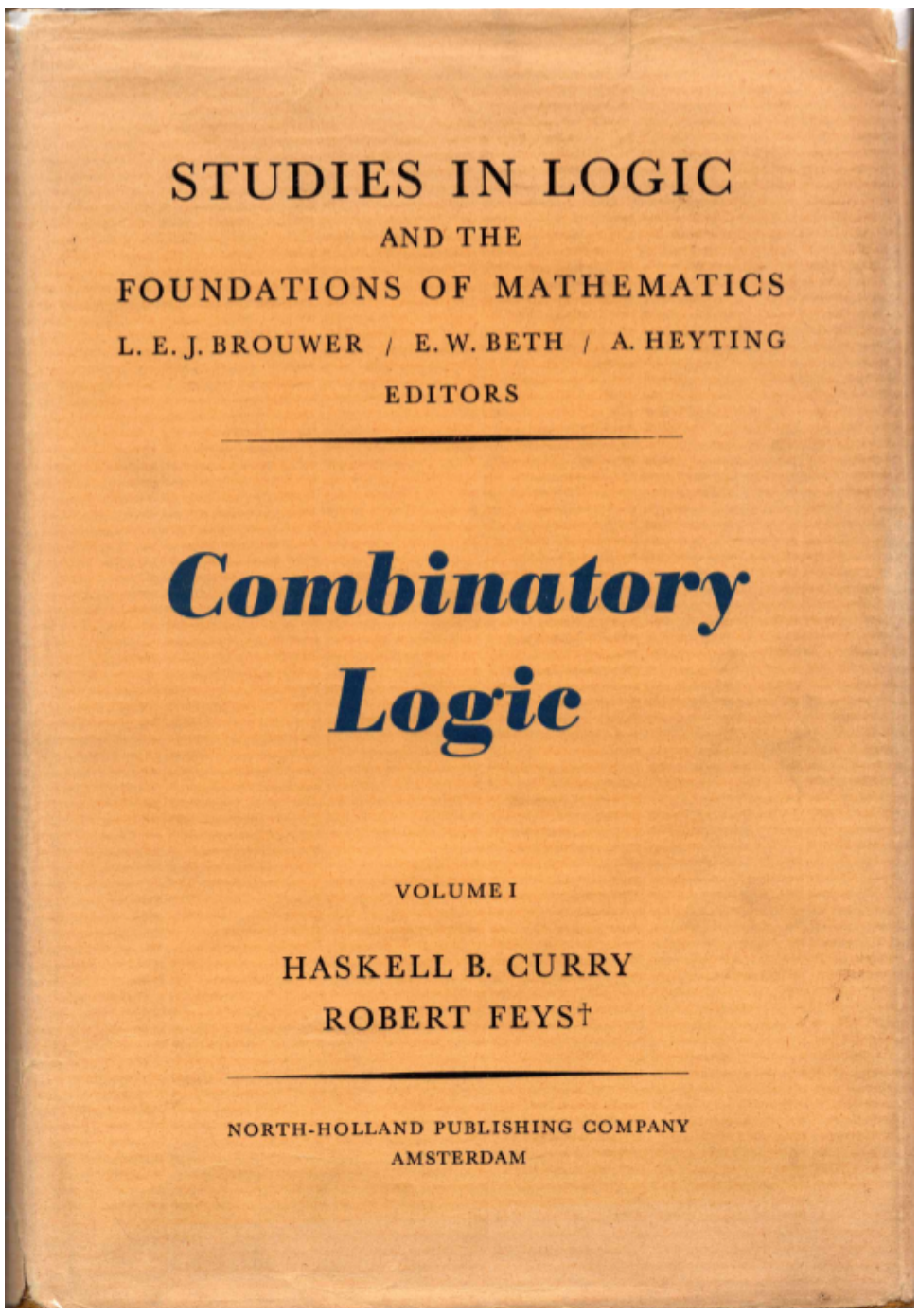

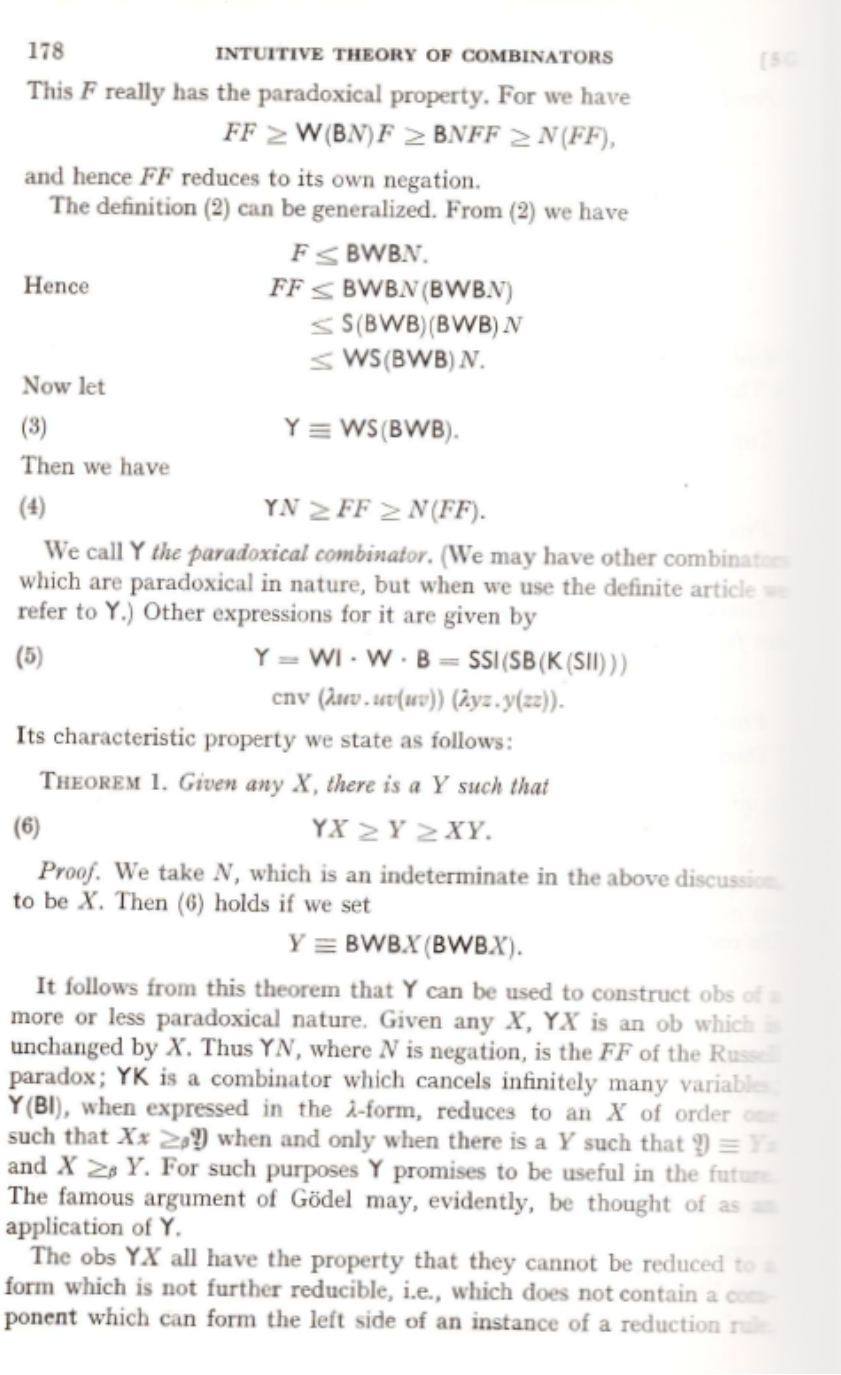

178 INTUITIVE THEORY OF COMBINATORS

This $F$ really has the paradoxical property. For we have

$$FF \geq \mathsf{W}(\mathsf{B}N)F \geq \mathsf{B}NFF \geq N(FF),$$

and hence $FF$ reduces to its own negation.

The definition (2) can be generalized. From (2) we have

$$F \leq \mathsf{BWB}N.$$

Hence

$$FF \leq \mathsf{BWB}N(\mathsf{BWB}N)$$
$$\leq \mathsf{S}(\mathsf{BWB})(\mathsf{BWB})N$$
$$\leq \mathsf{WS}(\mathsf{BWB})N.$$

Now let

(3) $$\mathsf{Y} \equiv \mathsf{WS}(\mathsf{BWB}).$$

Then we have

(4) $$\mathsf{Y}N \geq FF \geq N(FF).$$

We call Y *the paradoxical combinator*. (We may have other combinat which are paradoxical in nature, but when we use the definite article refer to Y.) Other expressions for it are given by

(5) $$\mathsf{Y} = \mathsf{WI} \cdot \mathsf{W} \cdot \mathsf{B} = \mathsf{SSI}(\mathsf{SB}(\mathsf{K}(\mathsf{SII})))$$
$$\text{cnv } (\lambda uv . uv(uv))\ (\lambda yz . y(zz)).$$

Its characteristic property we state as follows:

THEOREM 1. *Given any $X$, there is a $Y$ such that*

(6) $$\mathsf{Y}X \geq Y \geq XY.$$

*Proof.* We take $N$, which is an indeterminate in the above discuss to be $X$. Then (6) holds if we set

$$Y \equiv \mathsf{BWB}X(\mathsf{BWB}X).$$

It follows from this theorem that Y can be used to construct obs of more or less paradoxical nature. Given any $X$, $\mathsf{Y}X$ is an ob which unchanged by $X$. Thus $\mathsf{Y}N$, where $N$ is negation, is the $FF$ of the Russ paradox; YK is a combinator which cancels infinitely many variabl Y(BI), when expressed in the $\lambda$-form, reduces to an $X$ of order such that $Xx \geq_\beta \mathfrak{Y}$ when and only when there is a $Y$ such that $\mathfrak{Y} \equiv$ and $X \geq_\beta Y$. For such purposes Y promises to be useful in the futur The famous argument of Gödel may, evidently, be thought of as application of Y.

The obs $\mathsf{Y}X$ all have the property that they cannot be reduced to form which is not further reducible, i.e., which does not contain a com- ponent which can form the left side of an instance of a reduction

He called it the “paradoxical combinator” because it was recursively defined in a kind of self-referential way analogous to various paradoxes. Its explicit form is $SSK(S(K(SS(S(SSK))))K)$ and its most immediately notable feature is that under Schönfinkel’s combinator transformation rules it never settles down to a particular “value” but just keeps growing forever.

Well, in 2005 Paul Graham—who had long been an enthusiast of functional programming and LISP—decided to name his new (and now very famous) startup accelerator “Y Combinator”. I remember asking him why he’d called it that. “Because,” he said, “nobody understands the $Y$ combinator”.

Looking in my own archives from that time I find an email I sent a combinator enthusiast who was working with me:

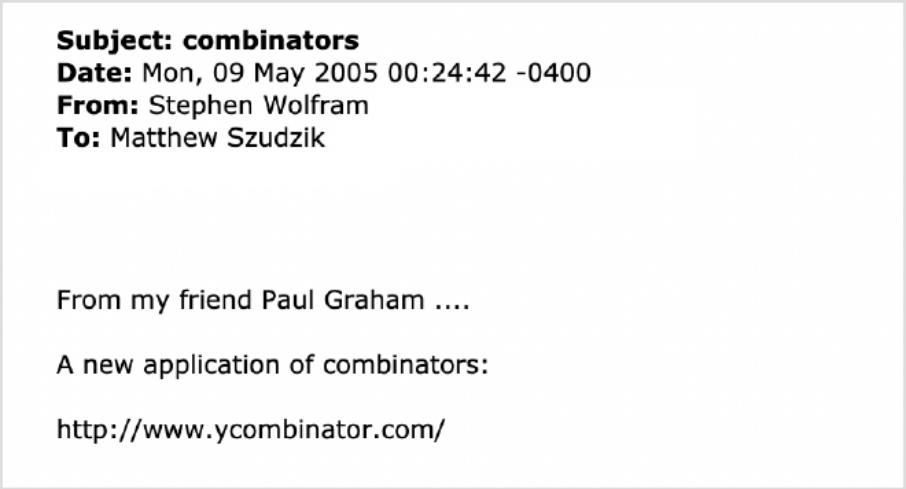

**Subject:** **combinators**
**Date:** Mon, 09 May 2005 00:24:42 -0400
**From:** Stephen Wolfram
**To:** Matthew Szudzik

From my friend Paul Graham ....

A new application of combinators:

http://www.ycombinator.com/

Followed by, basically, “Yes our theorem prover can prove the basic property of the $Y$ combinator” (V6 sounds so ancient; we’re now just about to release V12 .2):

```
(In a modern V6):

In[29]:=
FullSimplify[Exists[{Y},Y\[Equal]apply[combinator,Y]],
ForAll[{x,y},apply[apply[I,x],y]\[Equal]apply[x,apply[y,y]]]]

Out[29]=
True
```

I had another unexpected encounter with combinators last year. I had been given a book that was once owned by Alan Turing, and in it I found a piece of paper—that I recognized as being covered with none other than lambdas and combinators (but that's not *the* *Y* combinator):

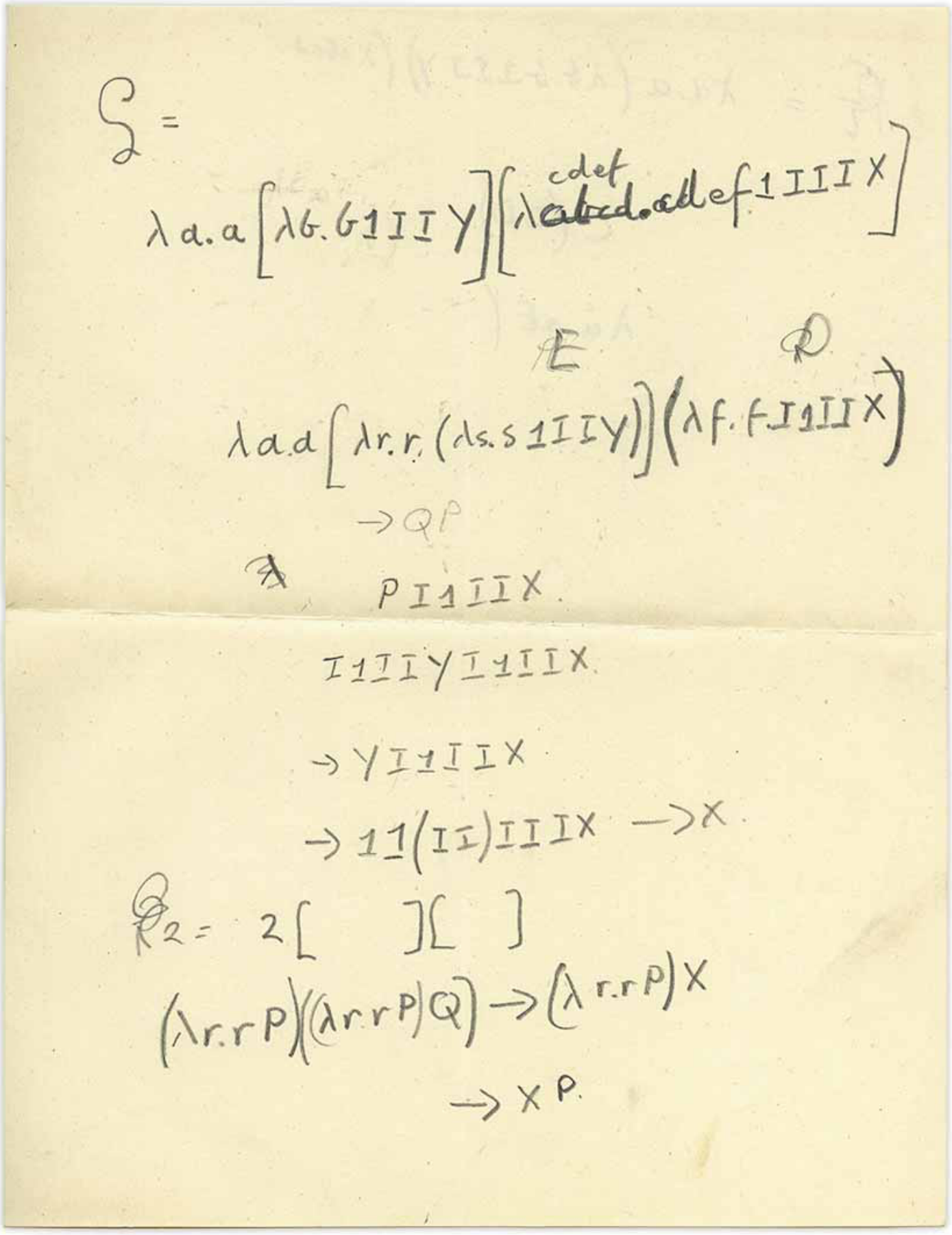

It took quite a bit of sleuthing (that I wrote extensively about)—but I eventually discovered that the piece of paper was written by Turing's student Robin Gandy. But I never figured out why he was doing combinators....

## Designing Symbolic Language

I think I first found out about combinators around 1979 by seeing Schönfinkel's original paper in a book called From *Frege to Gödel: A Source Book in Mathematical Logic* (by a certain Jean van Heijenoort). How Schönfinkel's paper ended up being in that book is an interesting question, which I'll write about elsewhere. The spine of my copy of the book has long been broken at the location of Schönfinkel's paper, and at different times I've come back to the paper, always thinking there was more to understand about it.

But why was I even studying things like this back in 1979? I guess in retrospect I can say I was engaged in an activity that goes back to Frege or even Leibniz: I was trying to find a fundamental framework for representing mathematics and beyond. But my goal wasn't a philosophical one; it was a very practical one: I was trying to build a computer language that could do general computations in mathematics and beyond.

My immediate applications were in physics, and it was from physics that my main methodological experience came. And the result was that—like trying to understand the world in terms of elementary particles—I wanted to understand computation in terms of its most fundamental elements. But I also had lots of practical experience in using computers to do mathematical computation. And I soon developed a theory about how I thought computation could fundamentally be done.

It started from the practical issue of transformations on algebraic expressions (turn $\sin(2x)$ into $2\sin(x)\cos(x)$, etc.). But it soon became a general idea: compute by doing transformations on symbolic expressions. Was this going to work? I wanted to understand as fundamentally as possible what computation really was—and from that I was led to its history in mathematical logic. Much of what I saw in books and papers about mathematical logic I found abstruse and steeped in sometimes horrendous notational complexity. But what were these people really doing? It made it much easier that I had a definite theory, against which I could essentially do reductionist science. That stuff in *Principia Mathematica*? Those ideas about rewriting systems? Yup, I could see how to represent them as rules for transformations on symbolic expressions.

And so it was that I came to design SMP: "A Symbolic Manipulation Program"—all based on transformation rules for symbolic expressions. It was easy to represent mathematical relations ($x is a pattern variable that would now in the Wolfram Language be x_ on the left-hand side only) :

```
STr[3,2,1]:    Asin[$x]+Asin[$y] -> Asin[$x Sqrt[1-$y^2]+$y Sqrt[1-$x^2]]
STr[3,2,2]:    Asin[$x]-Asin[$y] -> Asin[$x Sqrt[1-$y^2]-$y Sqrt[1-$x^2]]
STr[3,2,3]:    Acos[$x]+Acos[$y] -> Acos[$x $y+Sqrt[(1-$x^2) (1-$y^2)]]
STr[3,2,4]:    Acos[$x]-Acos[$y] -> Acos[$x $y-Sqrt[(1-$x^2) (1-$y^2)]]
```

Or basic logic:

```
/* Idempotent laws */
$p | $p : $p
$p & $p : $p

/* Commutative and associative laws built in */

/* Distributive laws */
$p | ($$q & $r) : ($p | $$q) & ($p | $r)

/* Identity laws built in */

/* Complement laws */
~~$p : $p
$p | ~$p : 1
$p & ~$p : 0

/* DeMorgan's laws */
~($$p | $q) : (~$$p) & (~$q)
~($$p & $q) : (~$$p) | (~$q)

/* Reflexive law */
$p => $p : 1

/* Antisymmetric law */
($p => $q) & ($q => $p) : $p=$q

/* Transitive law */
($p => $q) & ($q => $r) : $p=>$r
```

Or, for that matter, predicate logic of the kind Schönfinkel wanted to capture:

```
Quant[$s] : $s
Quant[$$q,Quant[$$r]] : Quant[$$q,$$r]

/* DeMorgan's law */
~Quant[All[$x],$$q,$s] : Quant[Some[$x],~Quant[$$q,$s]]
~Quant[Some[$x],$$q,$s] : Quant[All[$x],~Quant[$$q,$s]]
```

And, yes, it could emulate a Turing machine (note the tape-as-transformation-rules representation that appears at the end):

```
#I[2]::   Tape[$i_=$i<5]:1
#O[2]:    1
#I[3]::   Spec[1,0]:{Right,1}
#O[3]:    {Right,1}
#I[4]::   Spec[1,1]:{0,2}
#O[4]:    {0,2}
#I[5]::   Spec[2,0]:{Right,2}
#O[5]:    {Right,2}
#I[6]::   Spec[2,1]:{Right,1}
#O[6]:    {Right,1}
#I[7]::   Start[1,1]
#O[7]:    {5,1}
#I[8]::   Tape
#O[8]:    {[3]: 0, [1]: 0, [$i _= (5 > $i)]: 1}
```

But the most important thing I realized is that it really worked to represent basically anything in terms of symbolic expressions, and transformation rules on them. Yes, it was quite often useful to think of “applying functions to things” (and SMP had its version of lambda, for example), but it was much more powerful to think about symbolic expressions as just “being there” (“*x* doesn’t have to have a value”)—like things in the world—with the language being able to define how things should transform.

In retrospect this all seems awfully like the core idea of combinators, but with one important exception: that instead of everything being built from “purely structural elements” with names like *S* and *K*, there was a whole collection of “primitive objects” that were intended to have direct understandable meanings (like **Plus**, **Times**, etc.). And indeed I saw a large part of my task in language design as being to think about computations one might want to do, and then try to “drill down” to find the “elementary particles”—or primitive objects—from which these computations might be built up.

Over time I’ve come to realize that doing this is less about what one can in principle use to construct computations, and more about making a bridge to the way humans think about things. It’s crucial that there’s an underlying structure—symbolic expressions—that can represent anything. But increasingly I’ve come to realize that what we need from a computational language is to have a way to encapsulate in precise computational form the kinds of things we humans think about—in a way that we humans can understand. And a crucial part of being able to do that is to leverage what has ultimately been at the core of making our whole intellectual development as a species possible: the idea of human language.

Human language has given us a way to talk symbolically about the world: to give symbolic names to things, and then to build things up using these. In designing a computational language the goal is to leverage this: to use what humans already know and understand, but be able to represent it in a precise computational way that is amenable to actual computation that can be done automatically by computer.

It's probably no coincidence that the tree structure of symbolic expressions that I have found to be such a successful foundation for computational language is a bit like an idealized version of the kind of tree structure (think parse trees or sentence diagramming) that one can view human language as following. There are other ways to set up universal computation, but this is the one that seems to fit most directly with our way of thinking about things.

And, yes, in the end all those symbolic expressions could be constructed like combinators from objects—like *S* and *K*—with no direct human meaning. But that would be like having a world without nouns—a world where there's no name for anything—and the representation of everything has to be built from scratch. But the crucial idea that's central to human language—and now to computational language—is to be able to have layers of abstraction, where one can name things and then refer to them just by name without having to think about how they're built up "inside".

In some sense one can see the goal of people like Frege—and Schönfinkel—as being to "reduce out" what exists in mathematics (or the world) and turn it into something like "pure logic". And the structural part of that is exactly what makes computational language possible. But in my conception of computational language the whole idea is to have content that relates to the world and the way we humans think about it.

And over the decades I've continually been amazed at just how strong and successful the idea of representing things in terms of symbolic expressions and transformations on them is. Underneath everything that's going on in the Wolfram Language—and in all the many systems that now use it—it's all ultimately just symbolic expressions being transformed according to particular rules, and reaching fixed points that represent results of computations, just like in those examples in Schönfinkel's original paper.

One important feature of Schönfinkel's setup is the idea that one doesn't just have "functions" like *f*[*x*], or even just nested functions, like *f*[*g*[*x*]]. Instead one can have constructs where instead of the "name of a function" (like *f*) one can have a whole complex symbolic structure. And while this was certainly possible in SMP, not too much was built around it. But when I came to start designing what's now the Wolfram Language in 1986, I made sure that the "head" (as I called it) of an expression could itself be an arbitrary expression.

And when Mathematica was first launched in 1988 I was charmed to see more than one person from mathematical logic immediately think of implementing combinators. Make the definitions:

```
s[x_][y_][z_] := x[z][y[z]]
k[x_][y_] := x
```

Then combinators "just work" (at least if they reach a fixed point):

s[s[k[s]][s[k[k]][s[k[s]][k]]]][s[k[s[s[k][k]]]][k]][a][b][c]

a[b[a][c]]

But what about the idea of "composite symbolic heads"? Already in SMP I'd used them to do simple things like represent derivatives (and in Wolfram Language f'[x] is Derivative[1][f][x]). But something that's been interesting to me to see is that as the decades have gone by, more and more gets done with "composite heads". Sometimes one thinks of them as some kind of nesting of operations, or nesting of modifiers to a symbolic object. But increasingly they end up being a way to represent "higher-order constructs"—in effect things that produce things that produce things etc. that eventually give a concrete object one wants.

I don't think most of us humans are particularly good at following this kind of chain of abstraction, at least without some kind of "guide rails". And it's been interesting for me to see over the years how we've been able to progressively build up guide rails for longer and longer chains of abstraction. First there were things like Function, Apply, Map. Then Nest, Fold, FixedPoint, MapThread. But only quite recently NestGraph, FoldPair, SubsetMap, etc. Even from the beginning there were direct "head manipulation" functions like Operate and Through. But unlike more "array-like" operations for list manipulation they've been slow to catch on.

In a sense combinators are an ultimate story of "symbolic head manipulation": everything can get applied to everything before it's applied to anything. And, yes, it's very hard to keep track of what's going on—which is why "named guide rails" are so important, and also why they're challenging to devise. But it seems as if, as we progressively evolve our understand-ing, we're slowly able to get a little further, in effect building towards the kind of structure and power that combinators—in their very non-human-relatable way—first showed us was possible a century ago.

## Combinators in the Computational Universe

Combinators were invented for a definite purpose: to provide building blocks, as Schön-finkel put it, for logic. It was the same kind of thing with other models of what we now know of as computation. All of them were "constructed for a purpose". But in the end computation—and programs—are abstract things, that can in principle be studied without reference to any particular purpose. One might have some particular reason to be looking at how fast programs of some kind can run, or what can be proved about them. But what about the analog of pure natural science: of studying what programs just "naturally do"?

At the beginning of the 1980s I got very interested in what one can think of as the "natural science of programs". My interest originally arose out of a question about ordinary natural science. One of the very noticeable features of the natural world is how much in it seems to us highly complex. But where does this complexity really come from? Through what kind of mechanism does nature produce it? I quickly realized that in trying to address that question, I needed as general a foundation for making models of things as possible. And for that I turned to programs, and began to study just what "programs in the wild" might do.

Ever since the time of Galileo and Newton mathematical equations had been the main way that people ultimately imagined making models of nature. And on the face of it—with their real numbers and continuous character—these seemed quite different from the usual setup for computation, with its discrete elements and discrete choices. But perhaps in part through my own experience in doing mathematics symbolically on computers, I didn't see a real conflict, and I began to think of programs as a kind of generalization of the traditional approach to modeling in science.

But what kind of programs might nature use? I decided to just start exploring all the possibilities: the whole "computational universe" of programs—starting with the simplest. I came up with a particularly simple setup involving a row of cells with values 0 or 1 updated in parallel based on the values of their neighbors. I soon learned that systems like this had actually been studied under the name "cellular automata" in the 1950s (particularly in 2D) as potential models of computation, though had fallen out of favor mainly through not having seemed very "human programmable".

My initial assumption was that with simple programs I'd only see simple behavior. But with my cellular automata it was very easy to do actual computer experiments, and to visualize the results. And though in many cases what I saw was simple behavior, I also saw something very surprising: that in some cases—even though the rules were very simple—the behavior that was generated could be immensely complex:

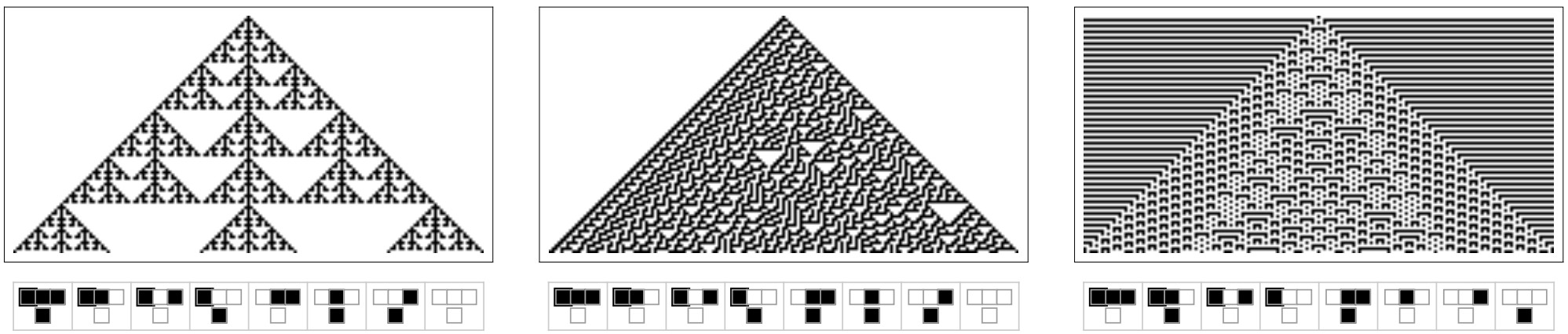

It took me years to come to terms with this phenomenon, and it's gradually informed the way I think about science, computation and many other things. At first I studied it almost exclusively in cellular automata. I made connections to actual systems in nature that cellular automata could model. I tried to understand what existing mathematical and other methods could say about what I'd seen. And slowly I began to formulate general ideas to explain what was going on—like computational irreducibility and the Principle of Computational Equivalence.

But at the beginning of the 1990s—now armed with what would become the Wolfram Language—I decided I should try to see just how the phenomenon I had found in cellular automata would play it in other kinds of computational systems. And my archives record that on April 4, 1992, I started looking at combinators.

I seem to have come back to them several times, but in a notebook from July 10, 1994 (which, yes, still runs just fine), there it is:

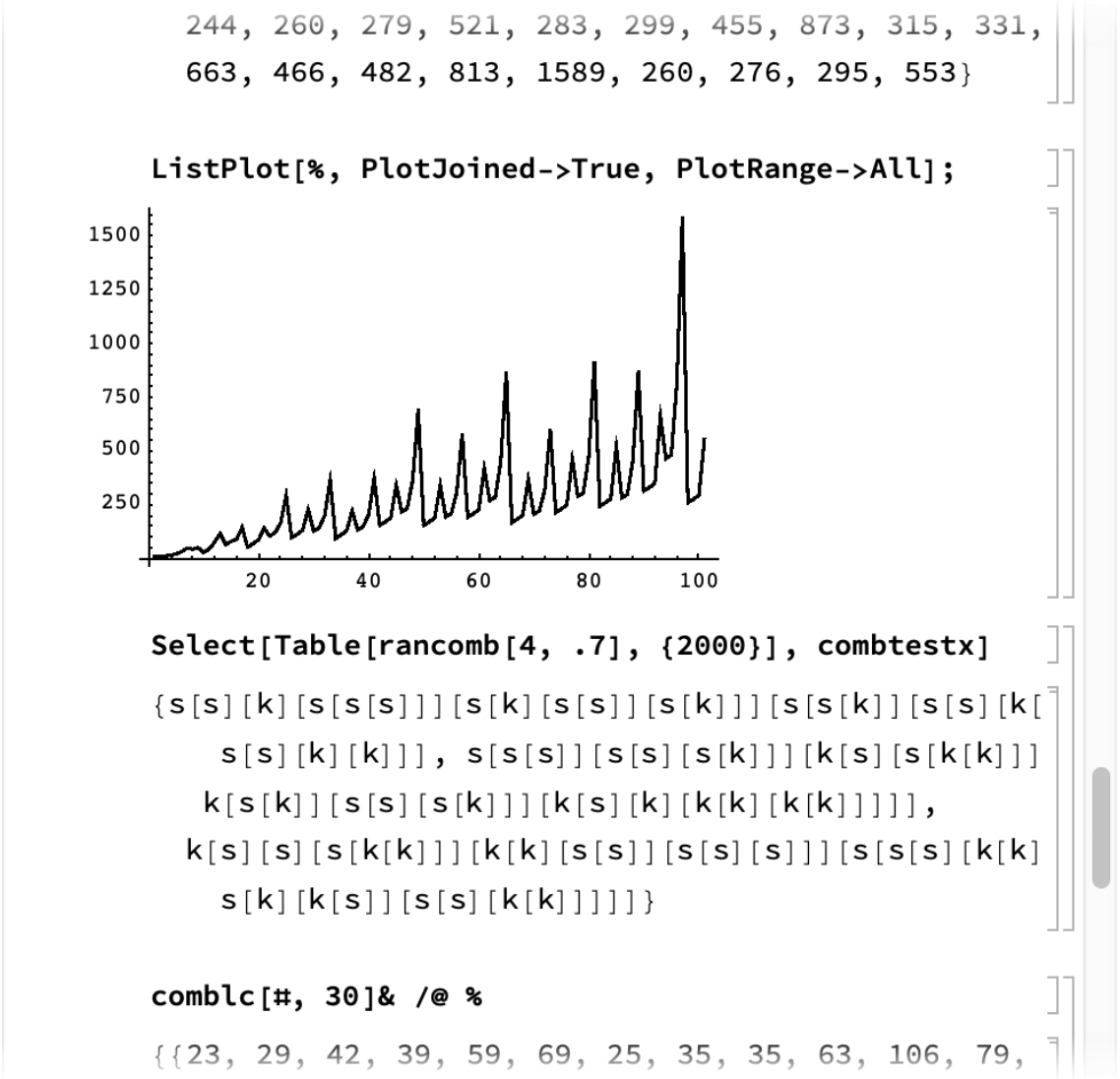

A randomly chosen combinator made of Schönfinkel's *S*'s and *K*'s starting to show complex behavior. I seem to have a lot of notebooks that start with the simple combinator definitions—and then start exploring:

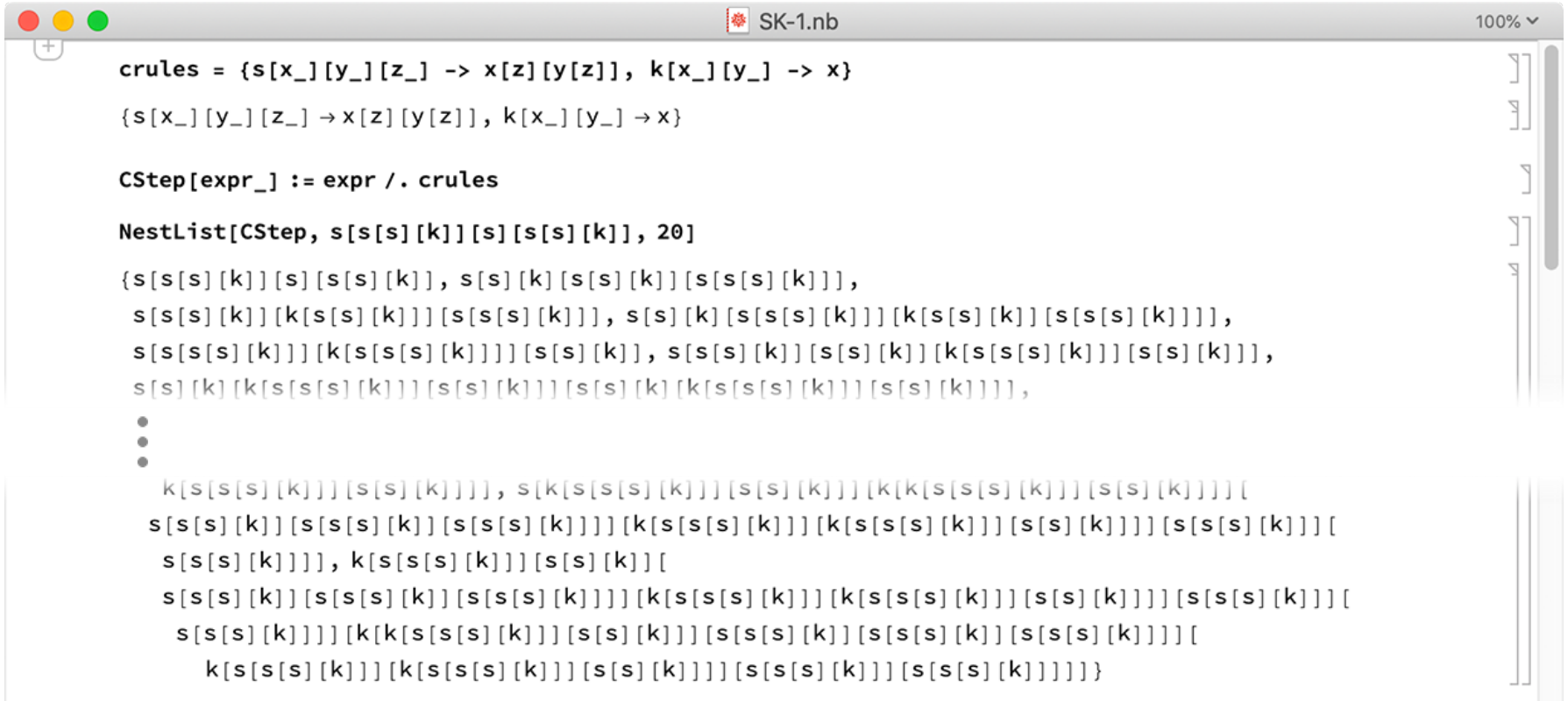

There are what seem like they could be pages from a "computational naturalist's field notebook":

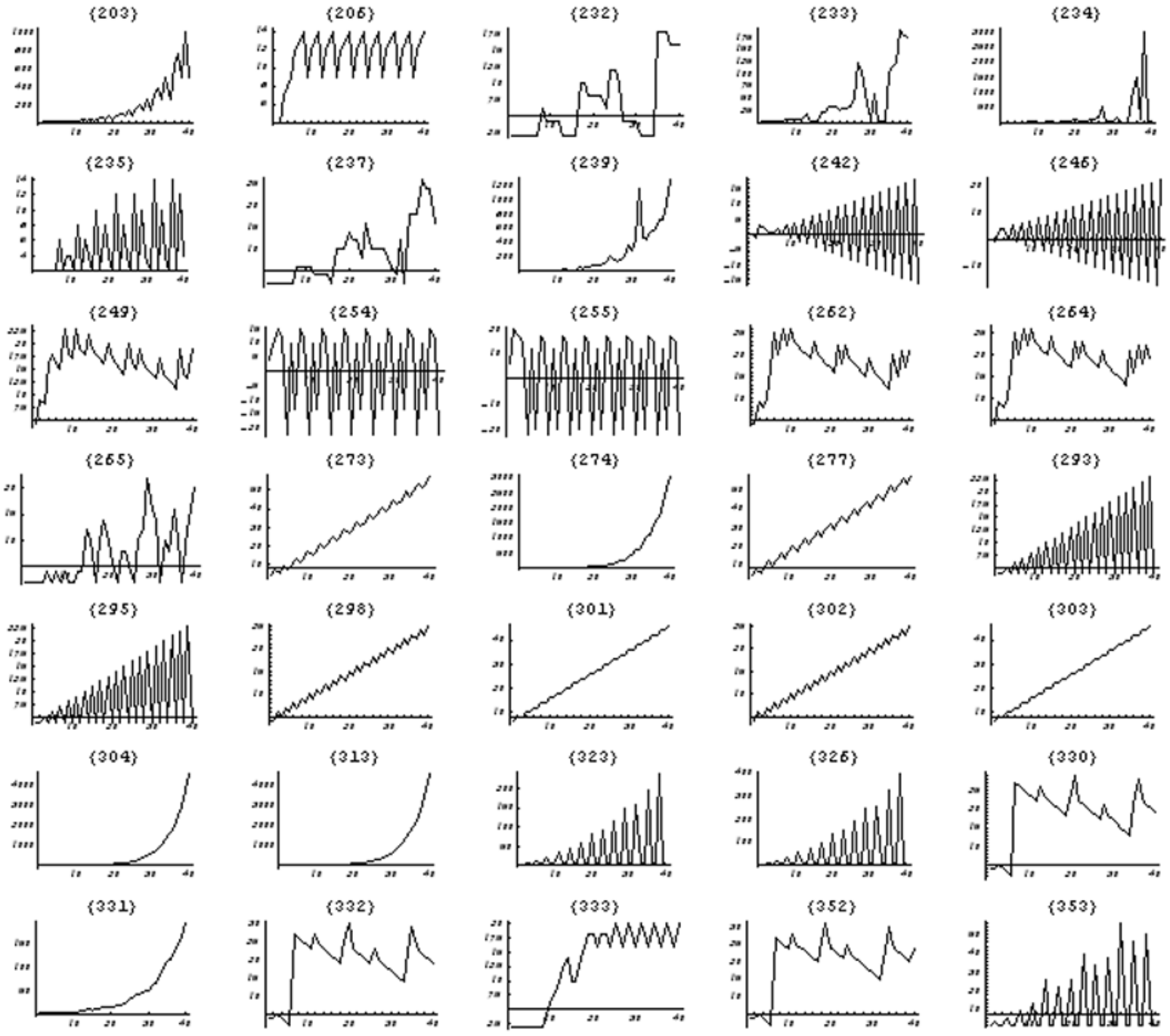

Then there are attempts to visualize combinators in the same kind of way as cellular automata:

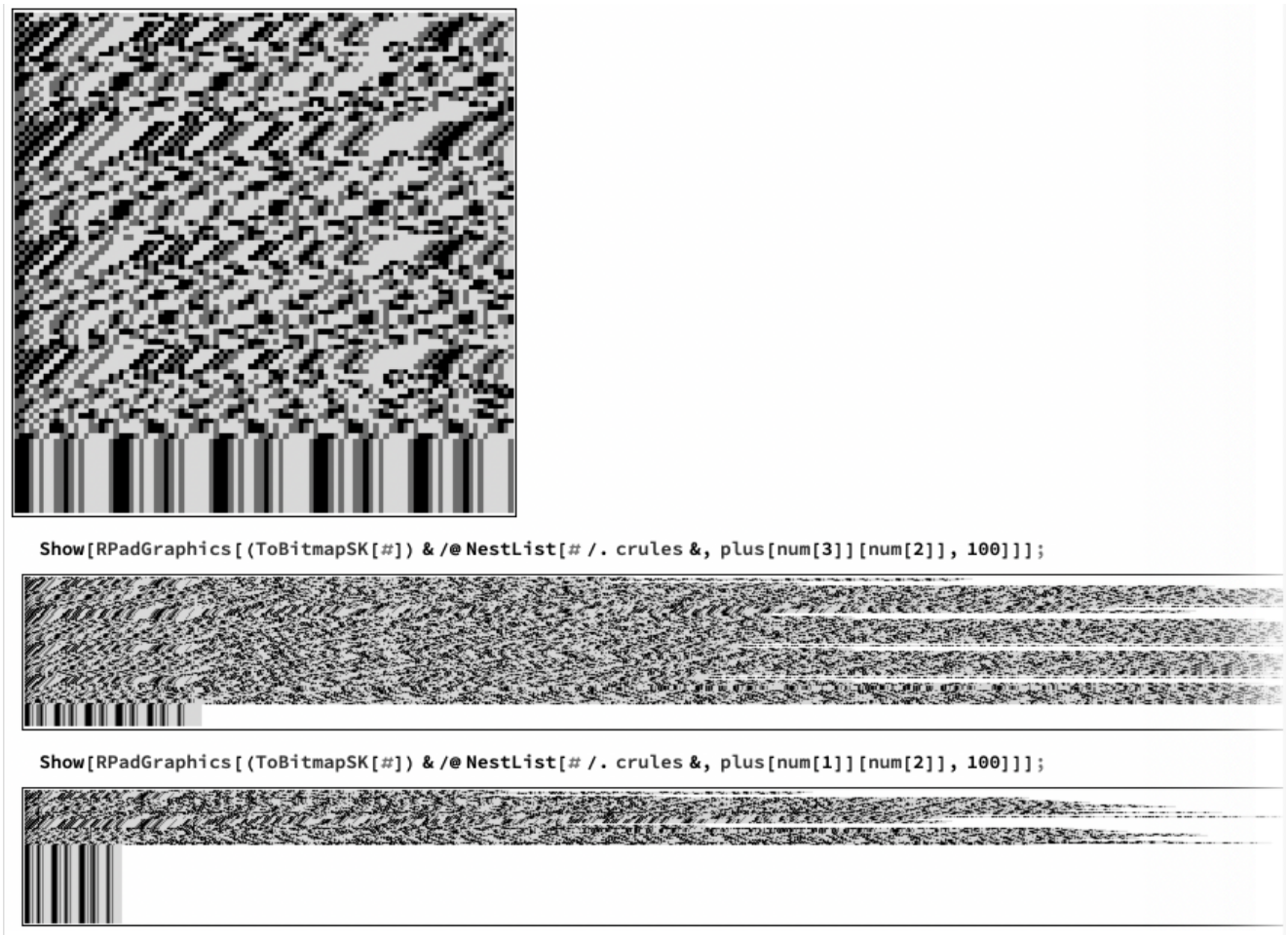

But the end result was that, yes, like Turing machines, string substitution systems and all the other systems I explored in the computational universe, combinators did exactly the same kinds of things I'd originally discovered in cellular automata. Combinators weren't just systems that could be set up to do things. Even "in the wild" they could spontaneously do very interesting and complex things.

I included a few pages on what I called "symbolic systems" (essentially lambdas) at the end of my chapter on "The World of Simple Programs" in *A New Kind of Science* (and, yes, reading particularly the notes again now, I realize there are still many more things to explore...):

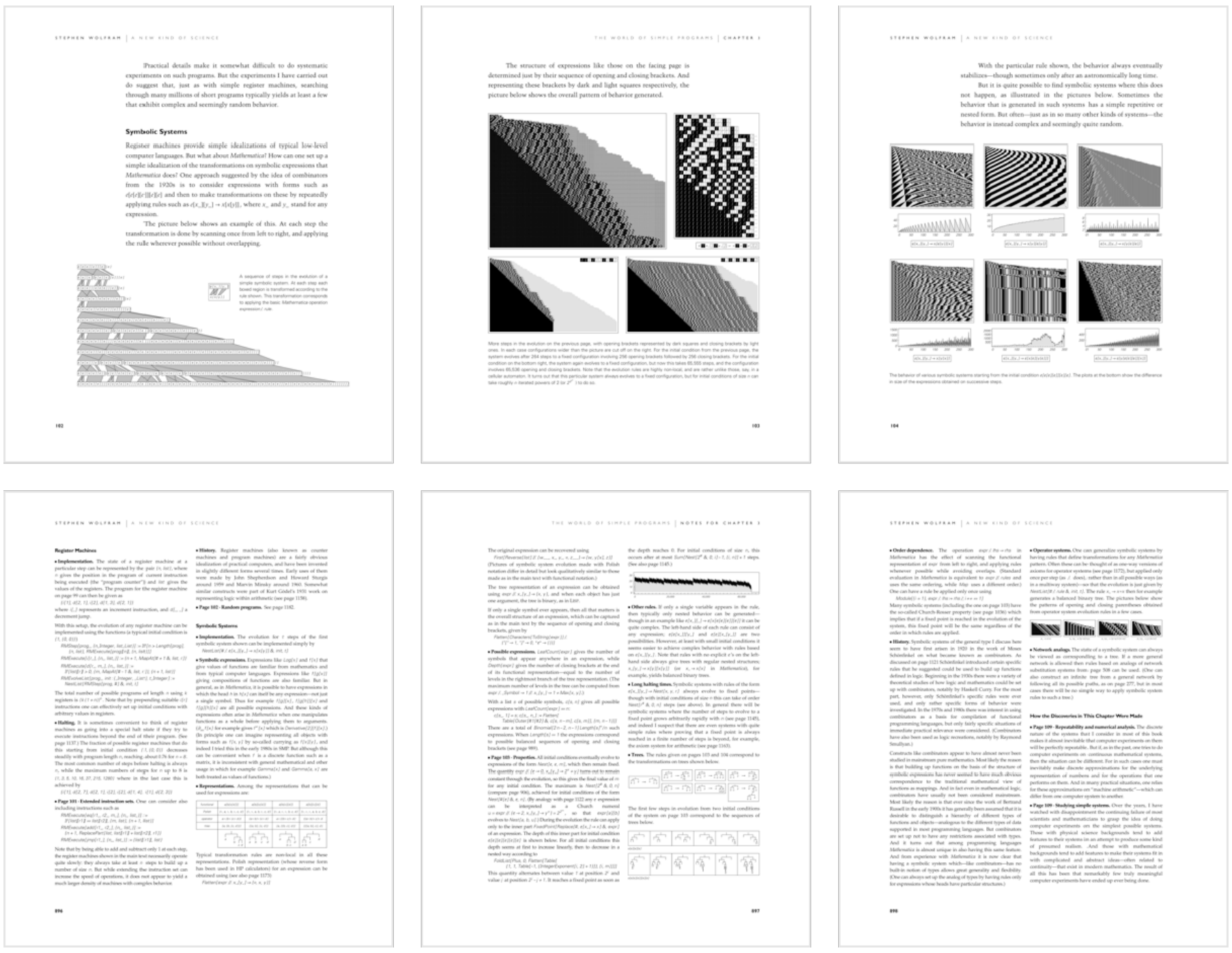

Later in the book I talk specifically about Schönfinkel's combinators in connection with the threshold of computation universality. But before showing examples of what they do, I remark:

"Originally intended as an idealized way to represent structures of functions defined in logic, combinators were actually first introduced in 1920—sixteen years before Turing machines. But although they have been investigated somewhat over the past eighty years, they have for the most part been viewed as rather obscure and irrelevant constructs."

How "irrelevant" should they be seen as being? Of course it depends on what for. As things to explore in the computational universe, cellular automata have the great advantage of allowing immediate visualization. With combinators it's a challenge to find any way to translate their behavior at all faithfully into something suitable for human perception. And since the Principle of Computational Equivalence implies that general computational features won't depend on the particulars of different systems, there's a tendency to feel that even in studying the computational universe, combinators "aren't worth the trouble".

Still, one thing that's been prominently on display with cellular automata over the past 20 or so years is the idea that any sufficiently simple system will eventually end up being a useful model for something. Mollusc pigmentation. Catalysis processes. Road traffic flow. There are simple cellular automaton models for all of these. What about combinators? Without good visualization it's harder to say "that looks like combinator behavior". And even after 100 years they're still a bit too unfamiliar. But when it comes to capturing some large-scale expression or tree behavior of some system, I won't be surprised if combinators are a good fit.

When one looks at the computational universe, one of the important ideas is "mining" it not just for programs that can serve as models for things, but also for programs that are somehow useful for some technological purpose. Yes, one can imagine specifically "compiling" some known program to combinators. But the question is whether "naturally occurring combinators" can somehow be identified as useful for some particular purpose. Could they deliver some new kind of distributed cryptographic protocol? Could they be helpful in mapping out distributed computing systems? Could they serve as a base for setting up molecular-scale computation, say with tree-like molecules? I don't know. But it will be interesting to find out. And as combinators enter their second century they provide a unique kind of "computational raw material" to mine from the computational universe.

## Combinators All the Way Down?

What is the universe fundamentally made of? For a long time the assumption was that it must be described by something fundamentally mathematical. And indeed right around the time combinators were being invented the two great theories of general relativity and quantum mechanics were just developing. And in fact it seemed as if both physics and mathematics were going so well that people like David Hilbert imagined that perhaps both might be completely solved—and that there might be a mathematics-like axiomatic basis for physics that could be "mechanically explored" as he imagined mathematics could be.

But it didn't work out that way. Gödel's theorem appeared to shatter the idea of a "complete mechanical exploration" of mathematics. And while there was immense technical progress in working out the consequences of general relativity and quantum mechanics little was discovered about what might lie underneath. Computers (including things like Mathematica) were certainly useful in exploring the existing theories of physics. But physics didn't show any particular signs of being "fundamentally computational", and indeed the existing theories seemed structurally not terribly compatible with computational processes.

But as I explored the computational universe and saw just what rich and complex behavior could arise even from very simple rules, I began to wonder whether maybe, far below the level of existing physics, the universe might be fundamentally computational. I began to make specific models in which space and time were formed from an evolving network of discrete points. And I realized that some of the ideas that had arisen in the study of things like combinators and lambda calculus from the 1930s and 1940s might have direct relevance.

Like combinators (or lambda calculus) my models had the feature that they allowed many possible paths of evolution. And like combinators (or lambda calculus) at least some of my models had the remarkable feature that in some sense it didn't matter what path one took;

the final result would always be the same. For combinators this "Church–Rosser" or "confluence" feature was what allowed one to have a definite fixed point that could be considered the result of a computation. In my models of the universe that doesn't just stop—things are a bit more subtle—but the generalization to what I call causal invariance is precisely what leads to relativistic invariance and the validity of general relativity.

For many years my work on fundamental physics languished—a victim of other priorities and the uphill effort of introducing new paradigms into a well-established field. But just over a year ago—with help from two very talented young physicists—I started again, with unexpectedly spectacular results.

I had never been quite satisfied with my idea of everything in the universe being represented as a particular kind of giant graph. But now I imagined that perhaps it was more like a giant symbolic expression, or, specifically, like an expression consisting of a huge collection of relations between elements—in effect, a certain kind of giant hypergraph. It was, in a way, a very combinator-like concept.

At a technical level, it's not the same as a general combinator expression: it's basically just a single layer, not a tree. And in fact that's what seems to allow the physical universe to consist of something that approximates uniform (manifold-like) space, rather than showing some kind of hierarchical tree-like structure everywhere.

But when it comes to the progression of the universe through time, it's basically just like the transformation of combinator expressions. And what's become clear is that the existence of different paths—and their ultimate equivalences—is exactly what's responsible not only for the phenomena of relativity, but also for quantum mechanics. And what's remarkable is that many of the concepts that were first discovered in the context of combinators and lambda calculus now directly inform the theory of physics. Normal forms (basically fixed points) are related to black holes where "time stops". Critical pair lemmas are related to measurement in quantum mechanics. And so on.

In practical computing, and in the creation of computational language, it was the addition of "meaningful names" to the raw structure of combinators that turned them into the powerful symbolic expressions we use. But in understanding the "data structure of the universe" we're in a sense going back to something much more like "raw combinators". Because now all those "atoms of space" that make up the universe don't have meaningful names; they're more like *S*'s and *K*'s in a giant combinator expression, distinct but yet all the same.

In the traditional, mathematical view of physics, there was always some sense that by "appropriately clever mathematics" it would be possible to "figure out what will happen" in any physical system. But once one imagines that physics is fundamentally computational, that's not what one can expect.

And just like combinators—with their capability for universal computation—can't in a sense be "cracked" using mathematics, so also that'll be true of the universe. And indeed in our model that's what the progress of time is about: it's the inexorable, irreducible process of computation, associated with the repeated transformation of the symbolic expression that represents the universe.

When Hilbert first imagined that physics could be reduced to mathematics he probably thought that meant that physics could be "solved". But with Gödel's theorem—which is a reflection of universal computation—it became clear that mathematics itself couldn't just be "solved". But now in effect we have a theory that "reduces physics to mathematics", and the result of the Gödel's theorem phenomenon is something very important in our universe: it's what leads to a meaningful notion of time.

Moses Schönfinkel imagined that with combinators he was finding "building blocks for logic". And perhaps the very simplicity of what he came up with makes it almost inevitable that it wasn't just about logic: it was something much more general. Something that can represent computations. Something that has the germ of how we can represent the "machine code" of the physical universe.

It took in a sense "humanizing" combinators to make them useful for things like computational language whose very purpose is to connect with humans. But there are other places where inevitably we're dealing with something more like large-scale "combinators in the raw". Physics is one of them. But there are others. In distributed computing. And perhaps in biology, in economics and in other places.

There are specific issues of whether one's dealing with trees (like combinators), or hypergraphs (like our model of physics), or something else. But what's important is that many of the ideas—particularly around what we call multiway systems—show up with combinators. And yes, combinators often aren't the easiest places for us humans to understand the ideas in. But the remarkable fact is that they exist in combinators—and that combinators are now a century old.

I'm not sure if there'll ever be a significant area where combinators alone will be the dominant force. But combinators have—for a century—had the essence of many important ideas. Maybe as such they are at some level destined forever to be footnotes. But in sense they are also seeds or roots—from which remarkable things have grown. And as combinators enter their second century it seems quite certain that there is still much more that will grow from them.

*Note: References are given as links in the body of this document.*